\documentclass[
               letterpaper,%
               twoside,%
               10pt%
               ]{article}

\usepackage[hmargin={1.0in,1.0in},
            vmargin={1.0in,1.0in},%
            headheight=14pt,
            ]{geometry}

\usepackage{amsmath}
\numberwithin{equation}{section}                        
\usepackage{amsthm}
\usepackage{amssymb}                                    
\usepackage{bbm}                                        
\makeatletter
\let\@upn\@iden
\makeatother
\newtheoremstyle{ToddTheorem}                           
  {3pt}
  {3pt}
  {\slshape}
  {}
  {\bfseries\upshape}
  {.}
  {.5em}
  {}

\newtheoremstyle{ToddDefinition}                        
  {3pt}
  {3pt}
  {\upshape}
  {}
  {\bfseries\slshape}
  {.}
  {.5em}
  {}

\makeatletter
\newcommand*{\c@theoremassumption}{\c@theorem}
\newcommand*{\p@theoremassumption}{\p@theorem}

\makeatother

\makeatletter
\newcommand*{\c@theoremconjecture}{\c@theorem}
\newcommand*{\p@theoremconjecture}{\p@theorem}

\makeatother

\makeatletter
\newcommand*{\c@theoremcorollary}{\c@theorem}
\newcommand*{\p@theoremcorollary}{\p@theorem}

\makeatother

\makeatletter
\newcommand*{\c@theoremdefinition}{\c@theorem}
\newcommand*{\p@theoremdefinition}{\p@theorem}

\makeatother

\makeatletter
\newcommand*{\c@theoremexample}{\c@theorem}
\newcommand*{\p@theoremexample}{\p@theorem}

\makeatother

\makeatletter
\newcommand*{\c@theoremfigure}{\c@theorem}
\newcommand*{\p@theoremfigure}{\p@theorem}

\makeatother

\makeatletter
\newcommand*{\c@theoremhypothesis}{\c@theorem}
\newcommand*{\p@theoremhypothesis}{\p@theorem}

\makeatother

\makeatletter
\newcommand*{\c@theoremlemma}{\c@theorem}
\newcommand*{\p@theoremlemma}{\p@theorem}

\makeatother

\makeatletter
\newcommand*{\c@theoremproposition}{\c@theorem}
\newcommand*{\p@theoremproposition}{\p@theorem}

\makeatother

\makeatletter
\newcommand*{\c@theoremremark}{\c@theorem}
\newcommand*{\p@theoremremark}{\p@theorem}

\makeatother

\makeatletter
\newcommand*{\c@theoremnotation}{\c@theorem}
\newcommand*{\p@theoremnotation}{\p@theorem}

\makeatother

\newtheoremstyle{SpringerTheorem}                           
  {3pt}
  {3pt}
  {\itshape}
  {}
  {\bfseries}
  {.}
  {.5em}
  {}
\newtheoremstyle{SpringerDefinition}                        
  {3pt}
  {3pt}
  {\rmfamily}
  {}
  {\bfseries}
  {.}
  {.5em}
  {}
\newtheoremstyle{SpringerExample}                        
  {3pt}
  {3pt}
  {\rmfamily}
  {}
  {\itshape}
  {.}
  {.5em}
  {}

\theoremstyle{SpringerTheorem} %

\theoremstyle{SpringerDefinition} %

\theoremstyle{SpringerExample} %

\newtheorem{remark}[theoremremark]{Remark}

\newcommand{\TheAuthor}{}
\newcommand{\Author}[1]%
        {\renewcommand{\TheAuthor}{#1}}                 
\newcommand{\TheRunningTitle}{}
\newcommand{\RunningTitle}[1]%
        {\renewcommand{\TheRunningTitle}{#1}}           

\usepackage{fancyhdr}                                   
\renewcommand{\footrulewidth}{0.4pt}

\usepackage{titling}
\pretitle{\begin{center}\LARGE\sffamily\slshape}        
\posttitle{\par\end{center}\vskip 0.5em}
\predate{\begin{center}\small}                          
\postdate{\par\end{center}\vskip 0.5em}
\thanksmarkseries{fnsymbol}                             

\usepackage[runin]{abstract}
\abslabeldelim{.\,}

\usepackage[titles]{tocloft}                            

\usepackage{titlesec}
\titleformat{\section}%
   {\Large\scshape\filcenter}{\thesection.}{1em}{}
\titleformat{\subsection}%
   {\large\sffamily\slshape\filcenter}{\thesubsection.}{1em}{}
\titleformat{\subsubsection}%
   {\normalsize\sffamily\filcenter}{\thesubsubsection.}{1em}{}
\titlespacing*{\section}%
   {0pt}{4.5ex plus 1ex minus .2ex}{3.3ex plus .2ex}
\titlespacing*{\subsection}%
   {0pt}{4.25ex plus 1ex minus .2ex}{3.05ex plus .2ex}
\titlespacing*{\subsubsection}%
   {0pt}{4.25ex plus 1ex minus .2ex}{3.05ex plus .2ex}

\usepackage[comma,numbers,square,sort&compress]{natbib} 

\usepackage{bm}                                         

\usepackage[
            ]{graphicx}                                 
\usepackage[
            small,%
            bf]{caption}                     
\setcaptionmargin{0.5in}

  {\begin{trivlist}\item[]\textbf{Acknowledgments.}}{\end{trivlist}}

\makeatletter
\providecommand*{\toclevel@assumption}{0}
\providecommand*{\toclevel@conjecture}{0}
\providecommand*{\toclevel@corollary}{0}
\providecommand*{\toclevel@definition}{0}
\providecommand*{\toclevel@example}{0}
\providecommand*{\toclevel@figure}{0}
\providecommand*{\toclevel@hypothesis}{0}
\providecommand*{\toclevel@lemma}{0}
\providecommand*{\toclevel@proof}{0}
\providecommand*{\toclevel@proposition}{0}
\providecommand*{\toclevel@remark}{0}
\providecommand*{\toclevel@theorem}{0}
\makeatother

\newcommand{\eref}[1]{\hyperref[{#1}]{(\ref*{#1})}}

\newcommand{\rmc}{\mathrm{c}}
\newcommand{\rmd}{\mathrm{d}}
\newcommand{\rme}{\mathrm{e}}

\newcommand{\rmi}{\mathrm{i}}

\newcommand{\rms}{\mathrm{s}}

\newcommand{\rmD}{\mathrm{D}}

\newcommand{\calL}{\mathcal{L}}

\newcommand{\vI}{\bm{\mathit{I}}}

\Author{T. Kapitula and P.G. Kevrekidis}
\RunningTitle{Language competition dynamics}

\usepackage[
            bookmarks=false,%
            bookmarksopen=false,%
            bookmarksnumbered=false,%
            hypertexnames=false,%
            colorlinks=true,%
            linkcolor=blue,%
            citecolor=blue,%
            urlcolor=blue]{hyperref}                    

\begin{document}

\title{Dynamics of two languages competing on a network: a case study}
\author{%
        Todd Kapitula %
        \thanks{E-mail: \href{mailto:tmk5@calvin.edu}{tmk5@calvin.edu}} \\
        Department of Mathematics and Statistics \\
        Calvin University \\
        Grand Rapids, MI 49546 %
\and
         Panayotis G. Kevrekidis %
         \thanks{E-mail: \href{mailto:kevrekid@math.umass.edu}{kevrekid@math.umass.edu}} \\
         Department of Mathematics and Statistics  \\
         University of Massachusetts \\
         Amherst, MA 01003-4515 %
        }

\begin{titlingpage}
\usethanksrule
\setcounter{page}{0}                                    
\maketitle
\begin{abstract}
A language dynamics model on a square lattice, which is an extension
of the one popularized by \citet{abrams:mtd03}, is analyzed using ODE
bifurcation theory. For this model we are interested in the existence
and spectral stability of structures such as stripes, which are realized through pulses
and/or the concatenation of fronts, and spots, which are a contiguous
collection of sites in which one language is dominant. Because the
coupling between sites is nonlinear, the boundary between sites
containing speaking two different languages is ``sharp''; in
particular, in a PDE approximation it allows for the existence of compactly supported pulses (compactons).
The dynamics are considered as a function of the prestige of a language. In particular, it is seen that as the prestige varies, it allows for a language to spread through the network, or conversely for its demise.
\end{abstract}

\vspace{2mm}


\cancelthanksrule
\renewcommand{\footrulewidth}{0.0pt}                    

\pdfbookmark[1]{\contentsname}{toc}                     
\tableofcontents                                        
\end{titlingpage}

\section{Introduction}

\begin{figure}[ht]
\begin{center}
\includegraphics{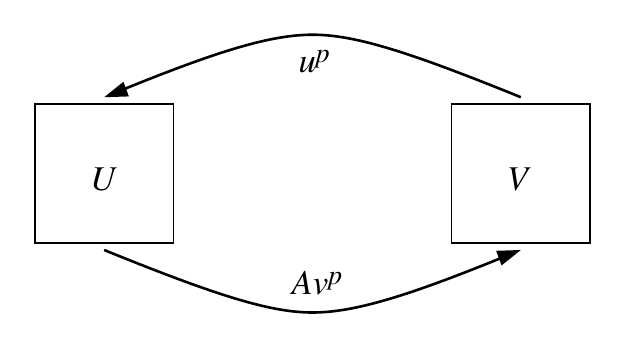}
\caption{(color online) The compartment model associated with AS model \eref{e:i1}. The variable $u$ represents the proportion of the population which speaks language $U$, and $v$ is the proportion which speaks language $V$. It is assumed $u+v=1$.}
\label{f:ASCompartmentModel}
\end{center}
\end{figure}

In their seminal paper \citet{abrams:mtd03} developed a simple ODE model,
\begin{equation}\label{e:i1}
	\dot{u}=(1-u)u^p-Au(1-u)^p,
\end{equation}
to help understand language competition and the decline in the number of people who speak such historic languages as Welsh, Quechua, and Scottish Gaelic. We will henceforth label \eref{e:i1} as the AS model (see \autoref{f:ASCompartmentModel} for a cartoon representation of this compartment model). The underlying assumptions for this model are that all speakers are monolingual, and the population is highly connected with no spatial or social structure. 
In equation \eref{e:i1} $u$ represents the proportion of the population which speak language $U$. If $v$ is the proportion which speak language $V$, since all speakers are monolingual, $v=1-u$. The parameter $p>0$ measures \textit{volatility}. The case $p=1$ is a neutral situation, where transition probabilities from one language to another depend linearly on local language densities. If $p>1$ there is a larger than neutral resistance to changing the language (low volatility), and if $p<1$ there is a lower than neutral resistance to changing the language (high volatility). Experimentally, it is estimated that $p=1.31\pm0.25$. The parameter $A>0$ represents the affinity of the general population towards one language or the other. In linguistics terminology the parameter $A$ can be used to represent the \textit{prestige} associated with a particular language. 

Assume $p>1$, so the volatility is low. The fixed points $u=0$ (language $V$ is preferred) and $u=1$ (language $U$ is preferred) are stable, while $u=B/(1+B)$ with $B=A^{1/(p-1)}$ is unstable. If $A<1$ and $u(0)=0.5$ (both languages are initially equally preferred), then $u(t)\to1$ as $t\to+\infty$, so the population has an affinity for language $U$. Or, language $U$ has more prestige in the general population. On the other hand, if $A>1$ and $u(0)=0.5$, then $u(t)\to0$ as $t\to+\infty$, so language $V$ has more prestige in the whole population. 

As pointed out by \citet{mira:isa05}, the monolingual assumption implies the two languages are so dissimilar that conversation is practically impossible between the two competing language groups. These authors extend the AS model to allow for languages which are similar enough for there to be bilingual speakers. The bilingual population subgroup satisfies, $b=1-u-v$, with $0\le b\le 1$. The model becomes,
\begin{equation}\label{e:i2}
\begin{split}
  \dot{u}&=(1-k)(1-u)(1-v)^p-Au(1-u)^p\\
  \dot{v}&=(1-k)A(1-v)(1-u)^p-v(1-v)^p,
  \end{split}
\end{equation}
where $0<k<1$ represents the ease of bilingualism. In particular, $k=0$ means that conversation is not possible between monolingual speakers, and $k=1$ implies $U=V$. The larger the value of $k$, the more similar are the two languages. If $k=b=0$, then model \eref{e:i2} reduces to model \eref{e:i1}. An analysis of the model \eref{e:i2} is provided in \citep{otero:aas13,colucci:cie14}.

An agent-based model associated with the AS model \eref{e:i1} when $p=1$,
\begin{equation}\label{e:i1a}
\dot{u}=(1-A)u(1-u),
\end{equation}
is considered by \citet{stauffer:mas07}. In particular, on a lattice of dimension $d$ an individual is assumed to feel the influence of $2d$ nearest neighbors. When $A>1$ the agent-based model results are qualitatively similar to those associated with the solution of \eref{e:i1a}. As expected, the results differ when the macroscopic model fails, $A=1$. It is not clear how the two models compare when $A<1$. An agent-based model is also considered by \citet{vazquez:abm10}.  In the fully connected case the dynamics of the associated mean-field model are equivalent to those for the model \eref{e:i1}.

The AS model has been extended to networks. Each node of the network corresponds to a group whose dynamics are governed by the AS model, and then the dynamics between groups satisfy some other rule. \citet{amano:gda14} collected and analyzed world-wide data taking into account such things as geographical range size, speaker population size, and speaker growth rate (i.e., changes in the number of speakers) of the world’s languages, and assessed interrelations among these three components to understand how they contribute to shaping extinction risk in languages. The role of population density and how it effects the interaction rates among groups is discussed by \citet{juane:uat19} in the context of language shift in Galicia, which is a bilingual community in northwest Spain.  They model the problem by looking at equations \eref{e:i2} on a network, with the strength of the interactions between nodes depending on the population density. The model for $j=1,\dots,n$ is,
\begin{equation}\label{e:i3}
\begin{split}
  \dot{u}_j&=(1-k_j)(1-u_j)(1-v_j)^p-A_ju_j(1-u_j)^p+K_j\left(\overline{u}-u_j\right)\\
  \dot{v}_j&=(1-k_j)A_j(1-v_j)(1-u_j)^p-v_j(1-v_j)^p+K_j\left(\overline{v}-v_j\right).
  \end{split}
\end{equation}
Here $\overline{y}$ represents the average of the set, $\{y_j\}$. The
positive parameter $K_j$ is assumed to be a strictly increasing
function of the population density. The authors \citet{franco:sme17}
follow a similar strategy, except they assume the
nonlinearities are of Lotka-Volterra type. Taking a different
approach, \citet{yun:tpo16} assume a diffusion process to take into
account spatial effects. \citet{fujie:amo13} and \citet{zhou:mce19} consider the problem of competition among more than two languages. 

In this paper we consider the language competition problem on a network under the assumption of low volatility, $p>1$. For ease we will primarily work with $p=2$, but our experience is that other values of $p>1$ do not effect the results qualitatively. We will assume there is no bilingual subpopulation (see \citep{juane:uat19,mira:tio11,otero:aas13} for some work in this area under the assumption of a single group). This may be an unrealistic assumption in terms of language; however, it is less so if one assumes language $U$ actually refers to those who have some type of religious affiliation, and language $V$ represents those who do not \citep{abrams:dos11}. We will assume the existence of $n$ distinct population groups, and let $0\le u_j\le 1$ represent the proportion of those in group $j$ who speak language $U$ ($v_j=1-u_j$ speak language $V$). For each $j=1,\dots,n$ our model equation is a natural extension of the compartment model illustrated in \autoref{f:ASCompartmentModel},
\begin{equation}\label{e:i4}
\dot{u}_j=\left(\sum_{k=1}^n\vI_{jk}u_k^p\right)\cdot(1-u_j)
-A_j\left(\sum_{k=1}^n\vI_{jk}(1-u_k)^p\right)\cdot u_j,
\end{equation}
where $\vI_{jk}\ge0$. We call the matrix $\vI=\left(\vI_{jk}\right)$ the influence matrix, and the term $\vI_{jk}$ represents the influence group $k$ has on group $j$ through the between-group reaction rate. If we think of the system \eref{e:i4} as being a compartment model, then the term $\vI_{jk}u_k^p$ is the rate constant associated with the influence that the $U$ speakers in group $k$ have on the $V$ speakers in group $j$, and $\vI_{jk}(1-u_k)^p$ is the rate constant associated with the influence that the $V$ speakers in group $k$ have on the $U$ speakers in group $j$. Clearly, when $n=1$ the system \eref{e:i4} collapses to the AS model \eref{e:i1}.

We now compare the systems \eref{e:i3} and \eref{e:i4}.
In the case of no bilingual speakers the system \eref{e:i3} collapses to,
\begin{equation}\label{e:i5}
  \dot{u}_j=(1-u_j)u_j^p-A_ju_j(1-u_j)^p+K_j\left(\overline{u}-u_j\right).
\end{equation}
The systems \eref{e:i4} and \eref{e:i5} have the feature that the on-site dynamics are the same as those for the AS model. However, the coupling between groups is different; in particular, the model \eref{e:i5} assumes that group $j$ is influenced by all of the other groups, whereas the model \eref{e:i4} allows for each group to be isolated from some of the other groups. Under the assumption that each external group has an equal influence on a given group, $\vI_{jj}=1$ and $\vI_{jk}=K_j/(n-1)$ for all $k\neq j$, the system \eref{e:i4} becomes,
\begin{equation}\label{e:i5aa}
\dot{u_j}=(1-u_j)\left[u_j^p+K_j\overline{u_{\neq j}^p}\right]-A_ju_j\left[(1-u_j)^p+
K_j\overline{(1-u)_{\neq j}^p}\right],
\end{equation}
where we use the notation,
\[
\overline{f_{\neq j}}=\frac1{n-1}\sum_{k\neq  j}f_k.
\]
The nonlinear coupling term for the model \eref{e:i5aa} is clearly very different than the linear coupling term associated with the model \eref{e:i5}. It is an open question as to whether this functional difference leads to a qualitative difference in the dynamics.

A simple model such as \eref{e:i1} can also be used to model opinion propagation in a population in which it is assumed that people have either opinion $U$, or opinion $V$, where we think of $V$ as being ``not $U$''. \citet{marvel:emc12}, hereafter referred to as MS, provide a model similar to \eref{e:i2} in which it is assumed there are three distinct groups: those who hold opinion $U$, those who hold opinion $V$, and the remaining who are undecided,
\[
\begin{split}
\dot{u}&=(1-u-v)u-uv\\
\dot{v}&=(1-u-v)v-uv.
\end{split}
\]
The underlying assumption in this model is that in order for one who initially holds opinion $U$ to eventually hold opinion $V$ (or vice-versa), the person first must become undecided. \citet{wang:bam16} extended the MS model to allow for several competing opinions. The MS model was extended to networks by \citet{bujalski:cac18}, and the extended model was studied using dynamical systems techniques. \citet{tanabe:cdo13} proposed and analyzed an interesting opinion formation model (hereafter labelled TM) in which it was assumed that the population itself breaks down into two groups: congregators, and contrarians. In contrast, the MS model implicitly assumes the entire population is filled with congregators. One conclusion of the TM model is that if a large enough proportion of the population is contrarian, then no majority opinion will be achieved. This is in contrast to the conclusion of those models in which it is assumed there are only congregators, as here a majority opinion is always obtained. The TM model was later refined by \citet{eekhoff:ofd19}, and the new model allowed for the effects of peer pressure, and incorporated the influence of zealots. From a qualitative perspective the mean-field models used for opinion
dynamics and language death have many similarities. Thus, although we
frame our results using the
formulation associated with language death, they are also directly applicable to mean-field opinion formation models.

In this paper we are primarily interested in the existence and stability of spatial structures for the network system \eref{e:i4}. We assume the groups have been arranged on a square lattice. The interactions on this lattice are nearest-neighbor (NN) only. Our experience is that from a qualitative perspective the NN interactions can be expanded without substantively changing the solution behavior as long as the interactions are still somewhat spatially localized  (the Implicit Function Theorem provides the theoretical justification). Moreover, there will be no preferential distinction in the reaction rates, $\vI_{jk}=\vI_{kj}$.  This is a case study, so we have not fully explored a large set of networks. That work will be left for a future paper. Our goal here is not to do an exhaustive study for all types of influence matrices. Instead, we simply want to get a sense of what is possible for a given type of network. 

For this lattice configuration we start by considering the existence and stability of fronts and pulses for the system \eref{e:i4}. A front is a solution for which $u_{jk}=U_j$, and $U_j=0$ (or $U_j=1$) for $1\le j\le n_0$, and $U_j=1$ (or $U_j=0$) for $j\ge n_0+\ell$ and some $\ell\ge1$. In other words, to the left of $n_0$ language $V$ is spoken, and to the right of $n_0+\ell$ language $U$ is spoken. A pulse is a solution for which $U_j=0$ for $j\le n_0$ and $j\ge n_0+\ell$, and $U_j>0$ for $n_0<j<n_0+\ell$. In other words, on the full lattice there is a stripe of language $U$ speakers who are surrounded by a group of $V$ speakers. We will consider when fronts can travel, which implies that language $U$ is invading language $V$, or vice-versa. We will also consider when pulses can grow or shrink. A growing pulse can be thought of as the concatenation of two fronts traveling in opposing directions, which implies that language $U$ eventually takes over the entire network. A shrinking pulse eventually disappears, which means that language $U$ has gone extinct. As we will see, the prestige associated with speaking $U\,(A<1)$ or $V\,(A>1)$ plays a central role in the analysis. We will conclude with a case study for a spot, which is a contiguous group of sites with $u_{jk}>0$ surrounded by $u_{jk}=0$ - an island of $U$ in a sea of $V$.

\vspace{2mm}\noindent\textbf{Acknowledgements.} %
This material is based upon work supported by the US National Science Foundation under Grant No. DMS-1809074 (PGK).

\section{The model on a square lattice}


As already stated, we consider the dynamics of a square lattice with
nearest-neighbor interactions only. Here $u_{jk}$ will represent the
proportion of the population at site $(j,k)$ who speak language
$U$. We will henceforth assume that the prestige associated with
language $U$ is uniform throughout the lattice, $A_{jk}=A$.
It is an interesting problem in its own right to allow for a spatially
inhomogeneous
distribution of the prestige and see how it affects the prevalent dynamics.
Moreover, we will assume $p=2$. Our numerical experiments indicate that from a qualitative perspective the results presented herein only need $p>1$. 

Under these assumptions the model \eref{e:i4}is,
\[
\begin{split}
\dot{u}_{jk}&=\left[\epsilon_0u_{jk}^2+
\epsilon_1\left(u_{j+1,k}^2+u_{j-1,k}^2+u_{j,k+1}^2+u_{j,k-1}^2\right)\right](1-u_{jk})\\
&\quad
-A\left[\epsilon_0(1-u_{jk})^2+
\epsilon_1\left((1-u_{j+1,k})^2+(1-u_{j-1,k})^2+(1-u_{j,k+1})^2+(1-u_{j,k-1})^2\right)\right]u_{jk}.
\end{split}
\]
Here $1\le j,k\le n$, and we assume in the model that at the edge of the square there are Neumann boundary conditions, e.g., $u_{n+1,k}=u_{nk}$. The parameter $\epsilon_0>0$ is the on-site interaction rate, and the parameter $\epsilon_1>0$ is the nearest-neighbor interaction rate.
Using the notation for the discrete Laplacian,
\[
\Delta_{\rmd\rmi\rms}f_{jk}=f_{j+1,k}+f_{j-1,k}+f_{j,k+1}+f_{j,k-1}-4f_{jk},
\]
the above ODE takes the more compact form,
\begin{equation}\label{e:2a2d}
\dot{u}_{jk}=\left(\epsilon_0+4\epsilon_1\right)u_{jk}(1-u_{jk})\left[(1+A)u_{jk}-A\right]\\
+2A\epsilon_1u_{jk}\Delta_{\rmd\rmi\rms}u_{jk}+
\epsilon_1\left[1-(1+A)u_{jk}\right]\Delta_{\rmd\rmi\rms}u_{jk}^2.
\end{equation}
If we assume that the interactions between neighbors are strong, i.e.,
$\epsilon_1\gg1$, then upon setting $R=\epsilon_0+4\epsilon_1\gg1$ we have the
limiting continuum model,
\begin{equation}\label{e:42d}
\partial_tu=R u(1-u)\left[(1+A)u-A\right]+(1+A)u(1-u)\Delta
u+\left[1-(1+A)u\right]\left|\nabla u\right|^2.
\end{equation}
Here $\Delta$ represents the Laplacian, and $\nabla$ is the gradient
operator.
The continuum model incorporates the expected temporal dynamics
associated with the original ODE model, but the coupling dynamics
between sites is dictated by an effective
nonlinear diffusion. The PDE is physical in the following sense: $u(x,y,t)=0$ implies $\partial_tu(x,y,t)\ge0$, and $u(x,y,t)=1$ implies $\partial_tu(x,y,t)\le0$. Note the diffusion coefficient vanishes when the entire population supports one language, $u=0$ or $u=1$.

When studying the solution structure to the ODE \eref{e:2a2d}, or the accompanying PDE \eref{e:42d}, we will first focus on the existence and spectral stability of time-independent patterns which vary in one direction only. For the ODE \eref{e:2a2d} we will set $u_{j,k}(t)=U_{j}$ for all $j,k$, and $U_j$ will solve the 1D discrete model,
\begin{equation}\label{e:2a1d}
0=\left(\epsilon_0+4\epsilon_1\right)U_j(1-U_j)\left[(1+A)U_j-A\right]+
2\epsilon_1AU_j\Delta_jU_j+
\epsilon_1\left[1-(1+A)U_j\right]\Delta_jU_j^2,
\end{equation}
where $\Delta_jf_j=f_{j+1}+f_{j-1}-2f_j$.
For the PDE \eref{e:42d} we will set $u(x,y,t)=U(x)$, and $U(x)$ will solve the nonlinear ODE,
\begin{equation}\label{e:42dode}
0=R U(1-U)\left[(1+A)U-A\right]+(1+A)U(1-U)U''+\left[1-(1+A)U\right](U')^2,\quad
 ^\prime=\frac{\rmd}{\rmd x}.
\end{equation}
In both cases we will be looking for fronts/pulses, which for the full system will correspond to stripes. These solutions act as transitions between regions where language $U$ is dominant and language $V$ is dominant.

\begin{remark}
	Even though the derivation is dissimilar, the continuum model
        \eref{e:42d} is remarkably similar to the mean-field model
        associated with the square lattice as provided for in
        \citep[equation~(48)]{vazquez:abm10}. The model \eref{e:42d}
        has the additional term, $\left[1-(1+A)u\right]\left|\nabla
          u\right|^2$; however, both models have the important feature
        that the diffusion coefficient is singular. Dynamically, both
        systems have the feature that small domains tend to shrink,
        and large domains tend to grow, and the domains tend to evolve
        in a way that reduces the curvature of the boundary; see also
        further relevant discussion regarding the dynamics below.
\end{remark}

\section{Existence and spectral  stability of stripes for the discrete model}

A front solution to \eref{e:2a1d} satisfies
$U_j=0\,(1)$ for $j\le\ell$, and $U_j=1\,(0)$ for $j\ge k$, where
$1<\ell<k<n$. A pulse solution will satisfy $U_j=0\,(1)$ for
$j\le\ell$ and $j\ge k$, and $U_j\sim1\,(0)$ for $\ell<j<k$.
The transition between the states 0 and 1 will be monotone.
A stripe solution to the full 2D model will be a pulse, or a
concatenation of two fronts. As we will see, the concatenation of two
fronts provides for a ``thicker'' stripe. In the same
spirit, we can also discuss multi-stripes, which are the concatenation of pulses and/or fronts.

\subsection{Existence: fronts}

If $\epsilon_1=0$,  the system uncouples, so a front can be
constructed analytically. In this limit, for a front we set $U_j=0\,(1)$ for
$j=1,\dots,\ell$, and $U_j=1\,(0)$ for $j=\ell+1,\dots,n$. We will refer to this front as the off-site front. Since each of the fixed points is stable for the scalar AS model, the front will be stable for the full system. By the Implicit Function Theorem the front will persist and be stable for $0<\epsilon_1\ll1$. We can concatenate these fronts when $\epsilon_1=0$ to form stable stripes, and then again apply the Implicit Function Theorem to show the existence and stability for small $\epsilon_1$.

When $\epsilon_1=0$ we can construct another front by setting $U_j=0\,(1)$ for $j=1,\dots,\ell,\,U_{\ell+1}=A/(1+A)$, and $U_j=1\,(0)$ for $j=\ell+2,\dots,n$. Since all of the fixed points but the one at $j=\ell$ are stable for the scalar AS model, the front will be unstable for the full system with the linearization having one positive eigenvalue. By the Implicit Function Theorem the front will persist and be unstable with one positive eigenvalue for $0<\epsilon_1\ll1$. We will refer to this front as the on-site front.

When $\epsilon_1=0$ the off-site and on-site fronts exist for any value of $A$. However, once there is nontrivial coupling, we expect there will be an interval of $A$ values which contains $A=1$ for which the fronts will exist. In order to determine this interval we will do numerical continuation using the MATLAB package, Matcont \citep{dhooge:mam03}. Using this package will also allow us to numerically continue bifurcation points in parameter space. Setting
\[
R_1=\frac{\epsilon_1}{\epsilon_0},
\]
we will numerically explore the $(R_1,A)$-parameter space. Since we analytically know what happens for $R_1=0$, we are in a good position to use numerical continuation.

\begin{figure}[ht]
\begin{center}
\begin{tabular}{cc}
\includegraphics{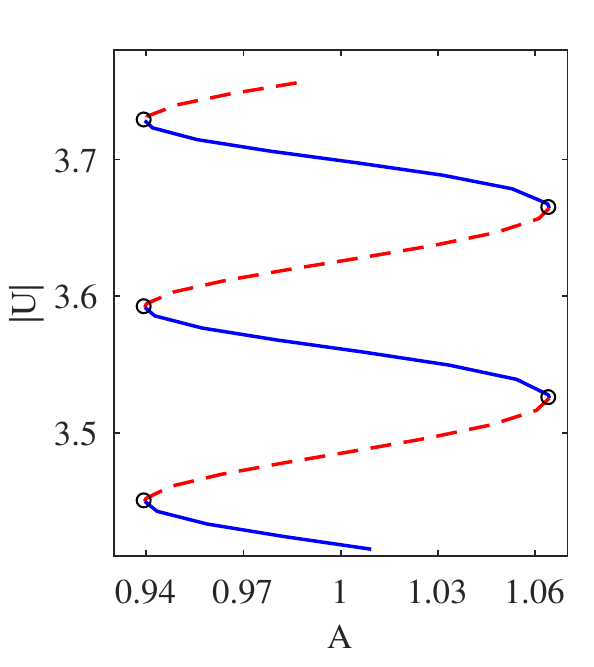}&
\includegraphics{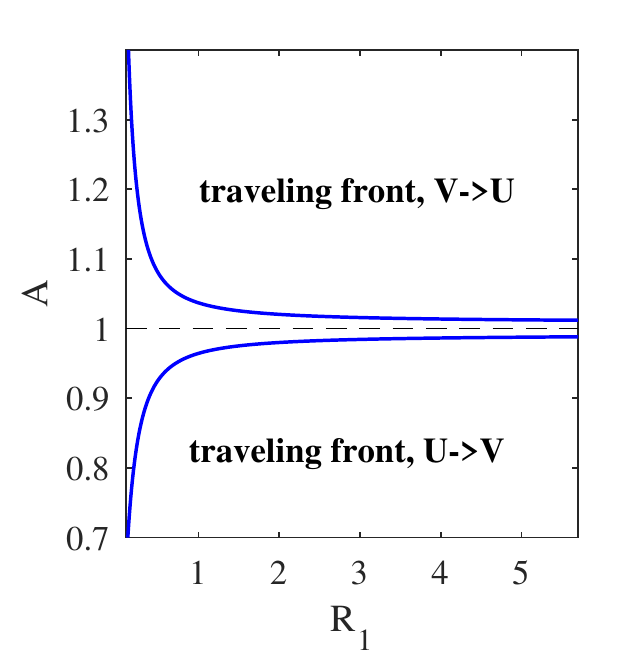}
\end{tabular}
\caption{(color online) Numerically generated existence curves for stationary $V\to U$ fronts, i.e., $U_j=0$ to the left, and $U_j=1$ to the right. The curve is given for $R_1=0.6$ in the left panel. The solid (blue) curves denote a stable front, and the dashed (red) curves denote an unstable front. The saddle-node bifurcation points are given by black circles. The vertical axis is the $\ell^2$-norm. Regarding the  boundary in the right panel, inside the two curves there is a stable stationary front, and outside the curves the front travels. The invading language is provided in the figure.}
\label{f:2DFront}
\end{center}
\end{figure}

For each fixed $R_1>0$ there will be an associated snaking diagram in
the parameter $A$. For a particular
example, consider the left figure in \autoref{f:2DFront}. The
horizontal axis is $A$, and the vertical axis is the $L^2$-norm of the
front. In this figure the solid (blue) curve corresponds to a stable
front (which is off-site when $A=1$), and the dashed (red) curve
corresponds to an unstable front (which is on-site when $A=1$). These
two curves meet at a saddle-node bifurcation point, which is denoted
by an open black circle. We see there is an $A_-<1<A_+$ for which
there are stable fronts for $A_-<A<A_+$, and no stationary fronts (at
least as seen via numerical continuation) outside this interval. The
values of $A_\pm$ depend on $R_1$
Each of the upward shifts of the stable and unstable branches correspond
to waveforms that are shifted by an integer number of lattice nodes to
the left (hence the growth in norm).
The right panel in\autoref{f:2DFront} shows the functions $A_\pm$ as a function of $R_1$. While we do not show it here, even in the limit $R_1\to+\infty$ the two curves do not converge to $1$; instead, we have $A_+(+\infty)\sim1.0082$, and $A_-(+\infty)\sim0.9918$. Inside the two curves, and for fixed $R_1$, there is a stable stationary front.

\begin{figure}[ht]
\begin{center}
\includegraphics{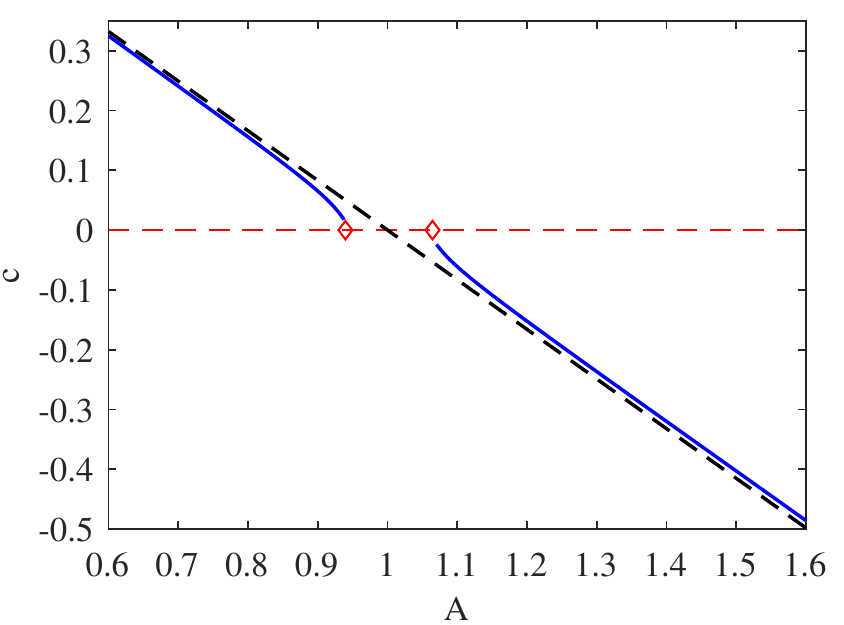}
\caption{(color online) The numerically generated wave speed when $R_1=0.6$ is given by the solid (blue) curve. The (red) diamonds mark the boundary for the existence of the stationary front, $A_-\sim0.9395$ and $A_+\sim1.0644$, at which the wave speed is zero. The dashed black line is the wave-speed prediction of \eref{e:cpred} provided by the PDE model.}
\label{f:2DFrontTravel}
\end{center}
\end{figure}

\subsection{Existence: traveling waves}\label{s:32aa}

Outside the two curves, $A_\pm(R_1)$, there is a traveling front. Traveling waves
will be written as $U(x+ct)$, so $U_j(t)=U(j+ct)$. Setting $\xi=x+ct$, the resulting forward-backward difference equation to which the traveling wave is a solution is,
\[
\begin{split}
cU'&=
\left(\epsilon_0+4\epsilon_1\right)U(1-U)\left[(1+A)U-A\right]\\
&\qquad+
2A\epsilon_1U\left[U(\xi+1)+U(\xi-1)-2U(\xi)\right]+
\epsilon_1\left[1-(1+A)U\right]\left[U(\xi+1)^2+U(\xi-1)^2-2U(\xi)^2\right].
\end{split}
\]
This system is solved using a variant of Newton's method (see \citep{hupkes:pfi11,hupkes:aon05,elmer:avo02} for the details).  

We consider in detail the case of $R_1=0.6$. Our experience is that
from a qualitative perspective the value of $R_1$ is not particularly
important. The numerical result is plotted in
\autoref{f:2DFrontTravel}. The points $A_\pm$ are marked with a (red)
diamond. It should be the case that at these points $c=0$;
unfortunately, the fact that the linearization becomes singular at
$A=A_\pm$ precludes good convergence of the algorithm near these
points. Away from these bifurcation points there is good convergence
of the numerical algorithm. Assuming $U_j=0$ to the left, and $V_j=0$ to the right, if $c<0$ language $V$ invades language $U$, whereas if $c>0$ language $U$ invades language $V$. We see here that if $A>A_+\sim1.0644$,
i.e., language $V$ has more prestige, then language $V$ invades
language $U$. On the other hand, if $A<A_-\sim0.9395$, i.e., language
$U$ has more prestige, then language $U$ invades language $V$. Note
that the speed increases as the preferred language becomes more
prestigious. Indeed, up to a small correction, and sufficiently far
away from $A_\pm$, the wave speed follows the formal prediction of the
continuum model, equation \eref{e:cpred}. The predicted curve, which
is associated with the limit $R_1\to+\infty$, is given by the black
dashed line. This result has been numerically verified for several
different values of $R_1$.
One can observe the nontrivial effect of discreteness in establishing
an interval where the fronts can be stationary. Indeed, the continuum
model is found to possess vanishing speed at the isolated point of
prestige balance, namely at $A=1$, while the discrete variant
requires a detuning from this value in order to enable such a
depinning from the vanishing speed setting.

\begin{remark}
	It is an interesting exercise to consider the scaling law for the wave speed as $A\to A_\pm$; however, we have not pursued this. The interested reader should consult \citet{anderson:pau16,kevrekidis:pfu01} and the references therein for details as to how such a law may be derived.
      \end{remark}

\subsection{Existence: pulses}\label{s:33aa}

As is the case for fronts, if $\epsilon_1=0$ a pulse can be constructed analytically by setting $U_j=0\,(1)$ for $1\le j\le\ell$ and $k\le j\le n$, and $U_j=1\,(0)$ for $\ell<j<k$. Since each of the fixed points is stable for the scalar AS model, the pulse will be stable for the full system. By the Implicit Function Theorem the pulse will persist and be stable for $0<\epsilon_1\ll1$. We can concatenate these pulses when $\epsilon_1=0$ to form stable stripes, and then again apply the Implicit Function Theorem to show the existence and stability for small $\epsilon_1$. If so desired, we can also construct unstable pulses by setting $U_{\ell+1}=A/(1+A)$ when $\epsilon_1=0$, and then using the Implicit Function Theorem for small $\epsilon_1$. Assuming the background supports language $V$, the size of the pulse is the number of adjacent groups which support language $U$. For small $\epsilon_1$ the size is $k-\ell-1$.

\begin{figure}[ht]
\begin{center}
\begin{tabular}{cc}
\includegraphics{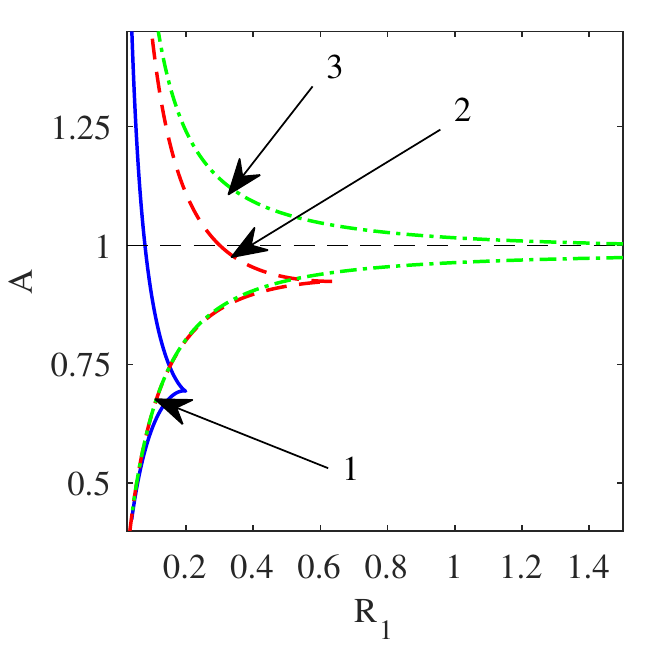}&
\includegraphics{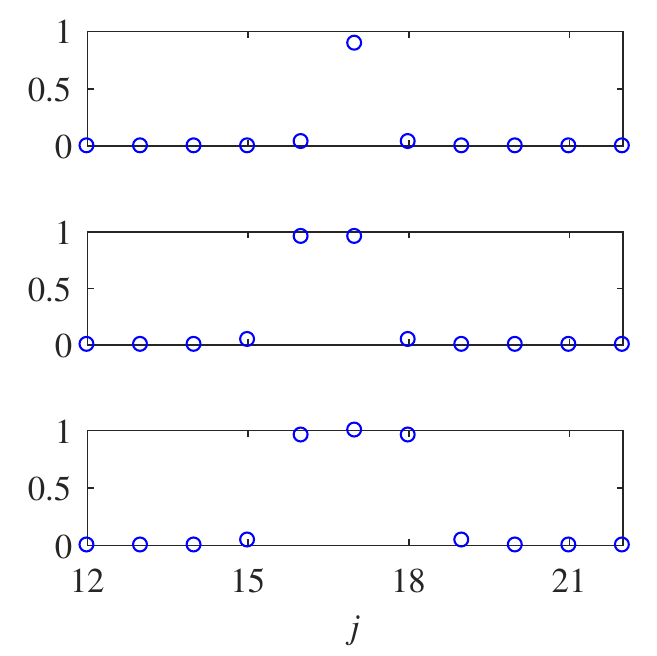}
\end{tabular}
\caption{(color online) The left panel provides the numerically
  generated boundary of pulses of size 1 through 3. The boundary is
  given by a solid (blue) curve for the pulse of size 1, a (red)
  dashed curve for a pulse of size 2, and a (green) dashed-dotted
  curve for a pulse of size 3. For a given pulse size, the pulse
  exists inside the two curves, and ceases to exist outside. The right panel gives an example of each pulse for
  $R_1=0.05$ and $A=1$. The pulse of size 1 is shown in the upper
  right panel, the pulse of size 2 in the middle right panel, and the
  pulse of size 3 in
  the lower right panel.}
\label{f:PulseLimitPoint}
\end{center}
\end{figure}

Numerically it is seen that if a pulse is of size 4 or larger, then it is realized as a concatenation of a $V\to U$ and a $U\to V$ stationary front. Consequently, the front dynamics completely determine the pulse dynamics. If the front is stationary, so is the pulse. If the front moves, so will the edge of the pulse. On the other hand, if the pulse is of size 1, 2, or 3, then the dynamics are not related to front dynamics. From a  dynamics perspective the pulse ceases to exist after a saddle-node bifurcation occurs.

Using Matcont, the bifurcation point can be traced in
$(R_1,A)$-space. The results are presented in
\autoref{f:PulseLimitPoint}. The pulse will exist inside the boundary
curve. The cusp point is $(R_1,A)\sim(0.1996,0.6936)$ for the pulse of
size 1, and $(R_1,A)\sim(0.6352,0.9246)$ for the pulse of size 2. For
a pulse of size 3 the cusp point satisfies $R_1>31.77$ with $0<1-A\ll1$, and is not
shown in the figure. Note that the cusp point converges to $A=1$ as the size of the pulse increases, and satisfies $A<1$. This is due to the fact that language $U$ has more prestige for $A<1$. If the background was language $U$ instead of language $V$, then the cusp point would satisfy $A>1$.

\begin{figure}[ht]
\begin{center}
\begin{tabular}{cc}
\includegraphics{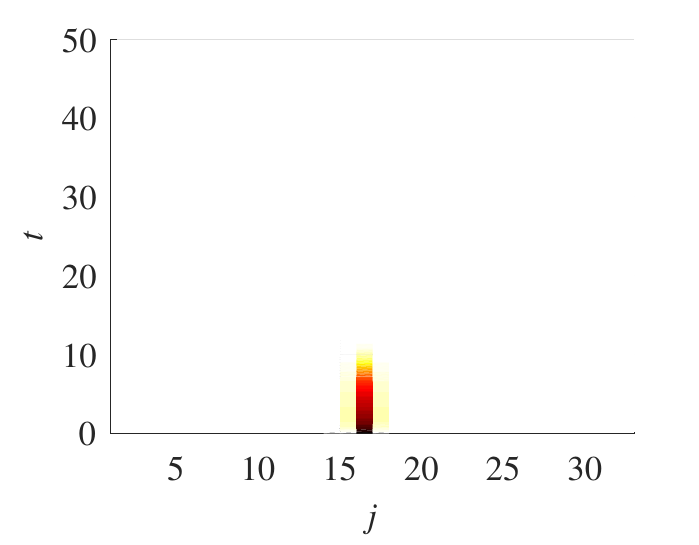}&
\includegraphics{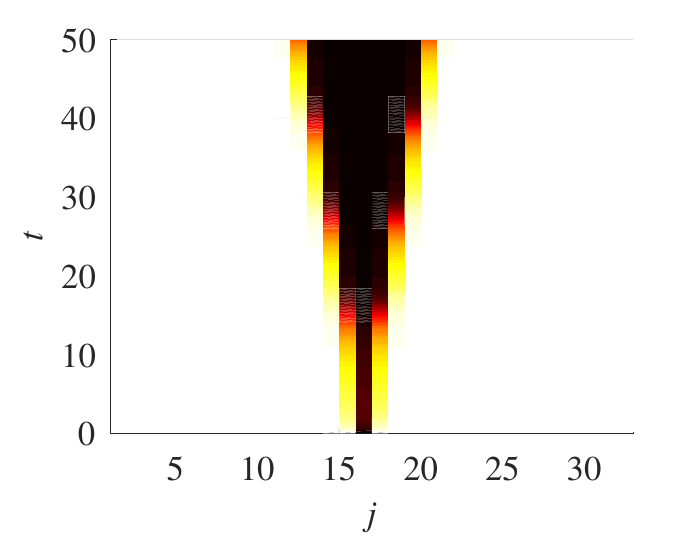}\\
\includegraphics{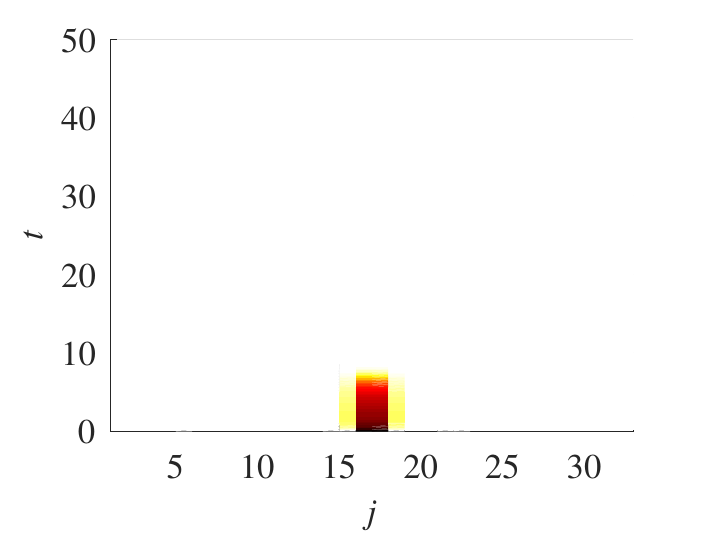}&
\includegraphics{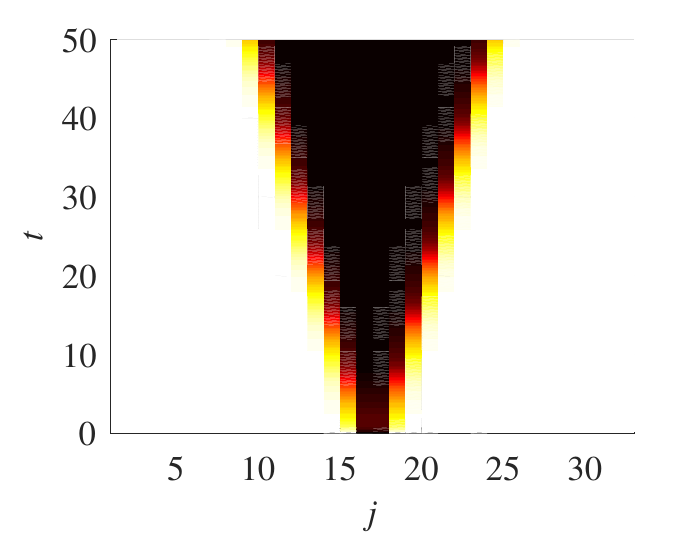}
\end{tabular}
\caption{(color online) The results of a numerical simulation of the full ODE \eref{e:2a2d} where the initial condition satisfies $u_{jk}(0)=u_{j\ell}(0)$ for all $k,\ell$. The color white represents language $V$, and the color black represents language $U$. In the top two figures $R_1=0.15$. For the top left figure $A=0.9$ (so the point is above the boundary for a pulse of size 1), and for the top right figure $A=0.6$ (so the point is below the boundary for a pulse of size 1). In both figures the initial condition for fixed $k$ is a small perturbation of a pulse of size 1. In the bottom two figures $R_1=0.5$. For the bottom left figure $A=1.0$ (so the point is above the boundary for a pulse of size 2), and for the bottom right figure $A=0.8$ (so the point is below the boundary for a pulse of size 2). In both figures the initial condition for fixed $k$ is a small perturbation of a pulse of size 2.}
\label{f:PulseGrowDecay}
\end{center}
\end{figure}

From a dynamics perspective, if $R_1$ is less than the cusp point value, and if $A$ is small enough so that $(R_1,A)$ is below the bottom boundary curve, then the pulse will grow until it can be thought of as a concatenation of two fronts. Once this occurs the edges of the pulse will move according to the front dynamics. The pulse grows because the prestige for language $U$ is sufficiently large. On the other hand, if $(AR_1,A)$ is above the top boundary curve, then language $V$ has sufficient prestige so that the background language prevails, and the pulse simply disappears in finite time. See \autoref{f:PulseGrowDecay} for the corroborating results of a particular simulation.

\begin{remark}
If we assume a pulse of language $V$ sits on a background of language $U$, then we will get the same curves as in \autoref{f:PulseLimitPoint}. However, the dynamical interpretation leading to \autoref{f:PulseGrowDecay} will be reversed. In particular, if $A$ is too small the pulse will disappear, whereas if $A$ is sufficiently large it will grow.
\end{remark}

\subsection{Multiple stripes via pulse concatenation}

\begin{figure}[ht]
\begin{center}
\includegraphics{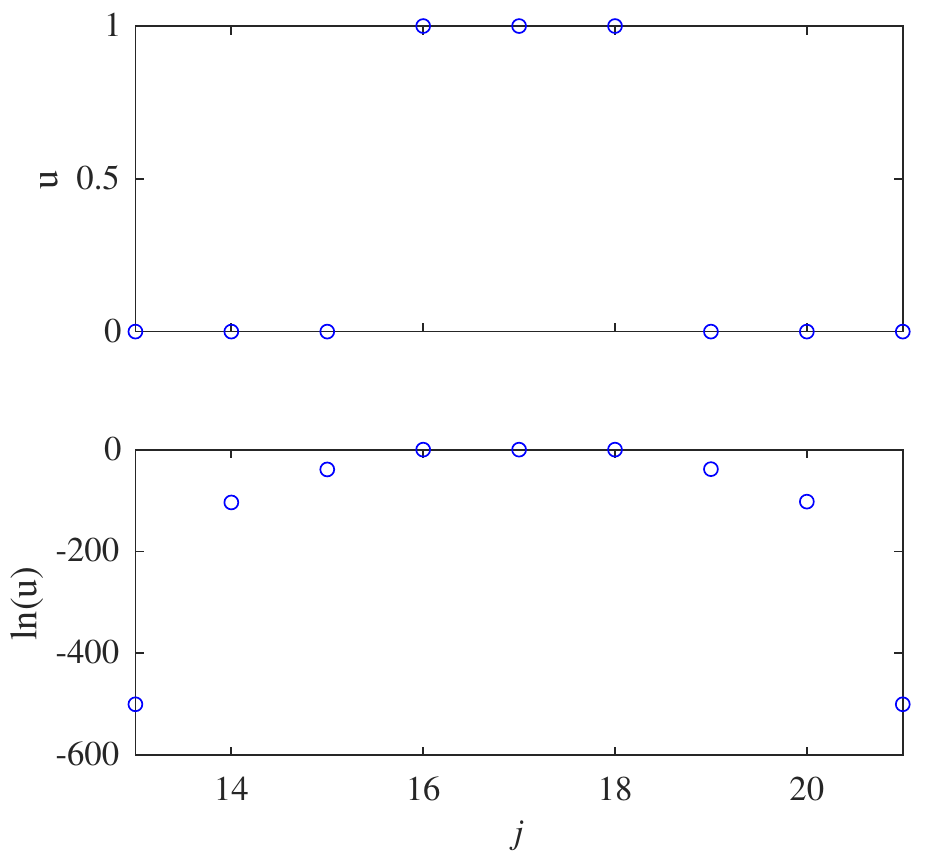}
\end{center}
\caption{(color online) The top panel provides the numerically generated pulse of size 3, say $u_3$, for $(R_1,A)=(0.9,1.0)$.  The bottom panel shows $\ln(u_3)$. For $j\le 12$ and $j\ge22$ the numerically determined value of $\ln(u_3)$ is $-\infty$. If the decay to $u=0$ was exponential, the bottom panel would be linear in $j$. Instead, it is concave down.}
\label{f:SuperExponentialDecay}
\end{figure}

We now consider the problem of concatenating individual pulses to form
multi-pulses. For the sake of convenience and without loss of
generality we assume that background consists of language $V$. As with the
single pulses, each of the multi-pulses will be stable when
$\epsilon_1=0$, and they will persist as stable structures for sufficiently
small $\epsilon_1$. Typically, the construction of multi-pulses would
involve a discussion of tail-tail interactions between individual pulses, and an application of the
Hale-Lin-Sandstede method (e.g., see
\citep{bramburger:lpi20,bramburger:sls20,sandstede:som98,hupkes:sop13,promislow:arm02,moore:rgr05,sandstede:sot02,parker:eas20}
and the references therein). However, for the system under consideration this is
less relevant, as the nonlinear coupling between adjacent sites renders
the transition from one state to another to be super-exponential, instead of
the exponential rates associated with linear coupling (see \autoref{f:SuperExponentialDecay} for a representative demonstration of this phenomena).
Consequently, to leading order one can think of pulses as being
compactons (a compactly supported structure), and fronts as being a
compactly supported transition between two states.\footnote{We will return to this aspect in more detail in the continuum limit
analysis, see \autoref{s:compacton}.}
In this light, to leading order, and as long as the individual pulses are initially sufficiently separated, the dynamics associated with a concatenation of $k$ pulses is really just the dynamics of $k$ uncoupled pulses, each of which evolves according to the rules presented in \autoref{s:33aa}.

Since this is only a case study, we will focus on the example of the two-pulse, which at the $\epsilon_1=0$ limit we  label as $j$-$k$-$\ell$. Here $j$ and $\ell$ refer to the size of the pulse which supports language $U$, and $k$ is the intervening pulse of size $k$ which supports language $V$. For example, a 2-1-2 can be thought of when $\epsilon_1=0$ as the sequence of $u$-values, $\cdots00\mathbf{11}0\mathbf{11}00\cdots$. 

\begin{figure}[ht]
	\begin{center}
		\begin{tabular}{cc}
			\includegraphics{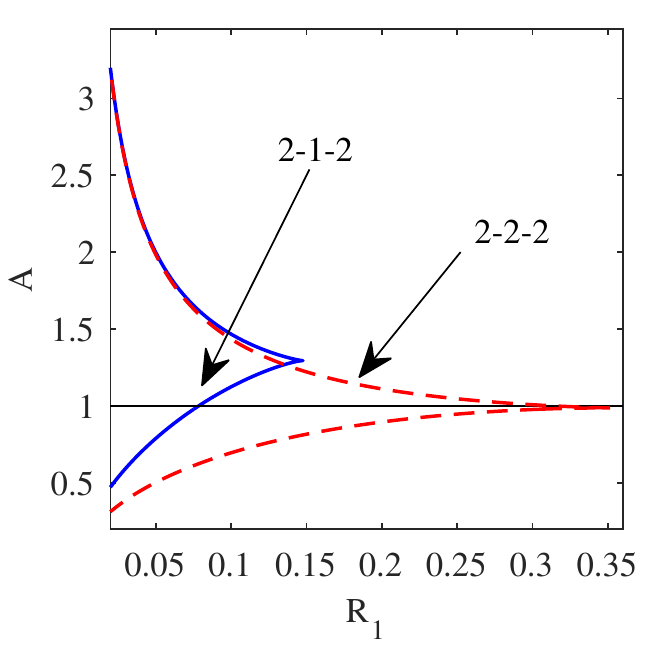}&
			\includegraphics{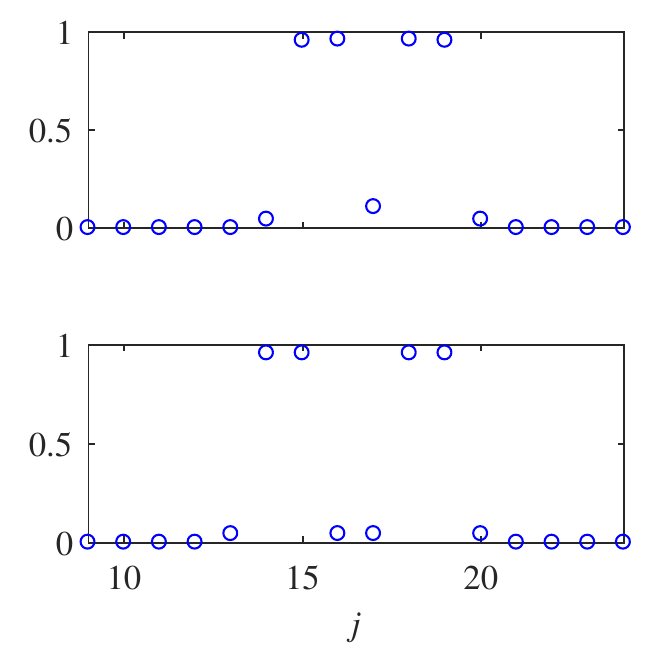}
		\end{tabular}
		\caption{(color online) The left panel provides the numerically generated boundary of the two-pulse 2-1-2 (solid (blue) curve) and 2-2-2 (dashed (red) curve). The two-pulse exists inside the two curves, and ceases to exist outside. The right panel gives an example of each pulse when $R_1=0.05$ and $A=1.0$. The 2-1-2 pulse is upper right, and the 2-2-2 pulse is lower right.}
		\label{f:MultiplePulseLimitPoint}
	\end{center}
\end{figure}

\begin{figure}[ht]
\begin{center}
\begin{tabular}{ccc}
\includegraphics{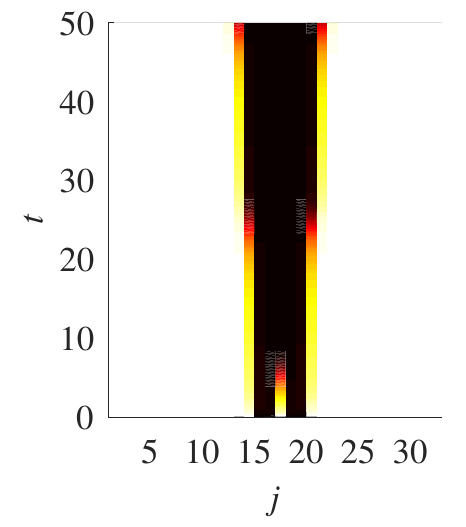}&
\includegraphics{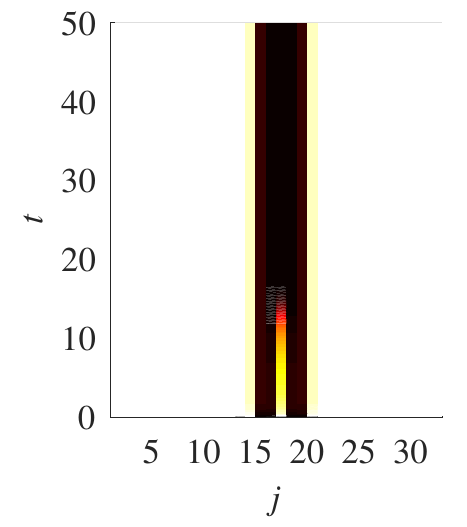}&
\includegraphics{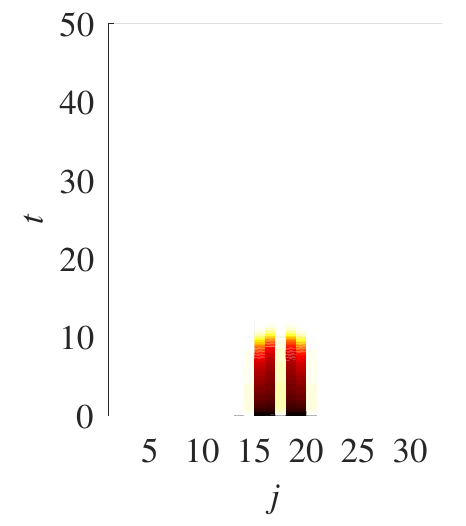}
\end{tabular}
\caption{(color online) The results of a numerical simulation of the full ODE \eref{e:2a2d} with $R_1=0.1$ where the initial condition satisfies $u_{jk}(0)=u_{j\ell}(0)$ for all $k,\ell$. The color white represents language $V$, and the color black represents language $U$. In all three panels the initial condition is a small perturbation of a 2-1-2 pulse. For the left panel $A=0.6$, for the middle panel $A=1.0$, and for the right panel $A=1.7$.}
\label{f:Size212}
\end{center}
\end{figure}

First consider the 2-1-2 pulse. The boundary for which this solution
exists is presented as a solid (blue) curve in
\autoref{f:MultiplePulseLimitPoint}. The cusp point is
$(R_1,A)\sim(0.1478,1.2954)$. For $(R_1,A)$ values inside the curve
the pulse will exist
as a stationary solution and be stable, whereas outside the curve it
does not exist. From a dynamical perspective, if $R_1<0.1478$, and $A$
is chosen so that the point lies below the lower boundary curve, then
the solution will quickly become a single pulse of size 5 (i.e., the
internal $0$ becomes a $1$), see the center panel of
\autoref{f:Size212} with $(R_1,A)=(0.1,1.0)$. As discussed previously,
a pulse of this size can be thought of as the concatenation of two
fronts. If the value of $A$ is such that the point is also below the
lower boundary of the curve presented in the right panel of
\autoref{f:2DFront}, so that $U$ invades $V$, then both fronts will
travel, i.e., expand until the entire lattice is overtaken by language
$U$ (see the left panel of \autoref{f:Size212} with
$(R_1,A)=(0.1,0.6)$). On the other hand, if $R_1<0.1478$, and $A$ is
chosen so that the point lies above the upper boundary curve, then the
solution will quickly decay to a pulse of size zero, i.e., language
$V$ is spoken over the entire lattice (see the right panel of
\autoref{f:Size212} with $(R_1,A)=(0.1,1.7)$). 

\begin{remark}
	If $R_1>0.1478$, then the pulse no longer exists, and the fate of the perturbation is a more difficult question to answer. This task will be left for a future paper.
\end{remark}

Next consider the 2-2-2 pulse. The boundary for which this solution exists is presented as a dashed (red) curve in \autoref{f:MultiplePulseLimitPoint}. The cusp point is $(R_1,A)\sim(0.3515,0.9879)$. The dynamics associated with $(R_1,A)$ points chosen outside of the domain bounded by the curve are exactly as that outlined above. For points below the curve the solution quickly becomes a single pulse of size 6, which again is the concatenation of two fronts. Each front will travel, and $U$ will grow, if $A$ is sufficiently small. For points above the curve the solution again quickly decays to a pulse of size zero.

Finally, consider the 2-$k$-2 pulse for any $k\ge3$. Here we find this is a true concatenation of two pulses of size 2, so the boundary curve is given by the dashed (red) curve in \autoref{f:PulseLimitPoint}. Moreover, the dynamics of this pulse is initially governed by the dynamics associated with a pulse of size 2 (see the bottom two panels of \autoref{f:PulseGrowDecay}).

While we do not present the corroborating details here, we now have the following rule-of-thumb. If we start with a two-pulse of size $j$-$k$-$\ell$, and if $k\ge3$, then the resulting dynamics will initially be independently governed by those associated with the pulse of size $j$ and pulse of size $\ell$. The individual pulses ``see'' each other only if the gap between the two is one or two adjacent sites. Indeed, this rule holds for any concatenation of pulses. As long as the distance between adjacent pulses is at least 3 sites, the existence boundary curve is exactly that associated with each individual pulse which makes up the entire multi-pulse. Moreover, the dynamics are governed by those associated with the single pulse until the distance between individual pulses is reduced to one or two sites.

\subsection{Spectral stability}

We have proven stable fronts and pulses exists for small $\epsilon_1$ for the 1D model \eref{e:2a2d}. We now remove the assumption that $\epsilon_1$ is small, and assume that a stable front/pulse exists for \eref{e:2a2d}. The spectrum for the associated linearized self-adjoint operator, $\calL_{1\rmD}$, is then strictly negative, so
\begin{equation}\label{e:2a11}
\langle\calL_{1\rmD}v_j,v_j\rangle<0.
\end{equation}

We now consider the spectral stability for the original 2D model \eref{e:2a2d}. The self-adjoint linearized operator has the form,
\[
\calL_{2\rmD}=\calL_{1\rmD}+2(1+A)\epsilon_1U_j(1-U_j)\Delta_k.
\]
Using a Fourier decomposition for the eigenfunctions in the transverse direction,
\[
v_{jk}\mapsto v_j\rme^{\rmi\xi k},\quad -\pi\le\xi<\pi,
\]
we find,
\[
\calL_{2\rmD}v_{jk}=\left[\calL_{1\rmD}
-4(1+A)\epsilon_1\left(1-\cos(\xi)\right)U_j(1-U_j)\right]v_j\rme^{\rmi\xi k}.
\]
Since the second term in the sum is a nonpositive operator, by using
the inequality \eref{e:2a11} we can conclude that
\[
\langle\calL_{2\rmD}v_{jk},v_{jk}\rangle<0.
\]
Consequently, all the eigenvalues must be strictly negative, so the stable front/pulse for the 1D problem is transversely stable for the 2D problem.

\section{Existence and spectral stability of stripes for the continuum model}

We now consider the existence and spectral stability of solutions to the continuum model \eref{e:42d}.

\subsection{Existence: compactons}\label{s:compacton}

The existence problem is settled by finding solutions to the nonlinear
ODE \eref{e:42dode}. Recalling $R=\epsilon_0+4\epsilon_1$, under the
assumption that neither language is more prestigious, $A=1$, there
exists the exact compacton solution,
\[
U_\rmc(x)=\frac12\left[1+\cos\left(\sqrt{\frac{R}2}\,x\right)\right].
\]
In writing this solutions there is the implicit understanding that the compacton is continuous with $U_\rmc(x)\equiv0$ or $U_\rmc(x)\equiv1$ outside some finite spatial interval. Of course, any spatial translation of the compacton is also a solution. Not only do these compactons define compactly supported pulses, they also define fronts connecting $u=0$ to $u=1$. One front satisfies $U_\rmc(x)=0$ for $x\le-\pi\sqrt{2/R}$, and $U_\rmc(x)=1$ for $x\ge0$ (of course, this front can be translated). Another front satisfies $U_\rmc(x)=1$ for $x\le0$, and $U_\rmc(x)=0$ for $x\ge\pi\sqrt{2/R}$ (again, this front can be translated). Note that the width of the front/pulse depends upon the reaction rate, $R$.

\begin{remark}
There is also an explicit compact solution when $p=3$,
\[
U_\rmc(x)=\frac12\left[1+\cos\left(\sqrt{\frac{2R}3}\,x\right)\right].
\]
Numerically, we see compactons for any $p>1$.
\end{remark}

\subsection{Traveling waves}

\begin{figure}[ht]
\begin{center}
\includegraphics{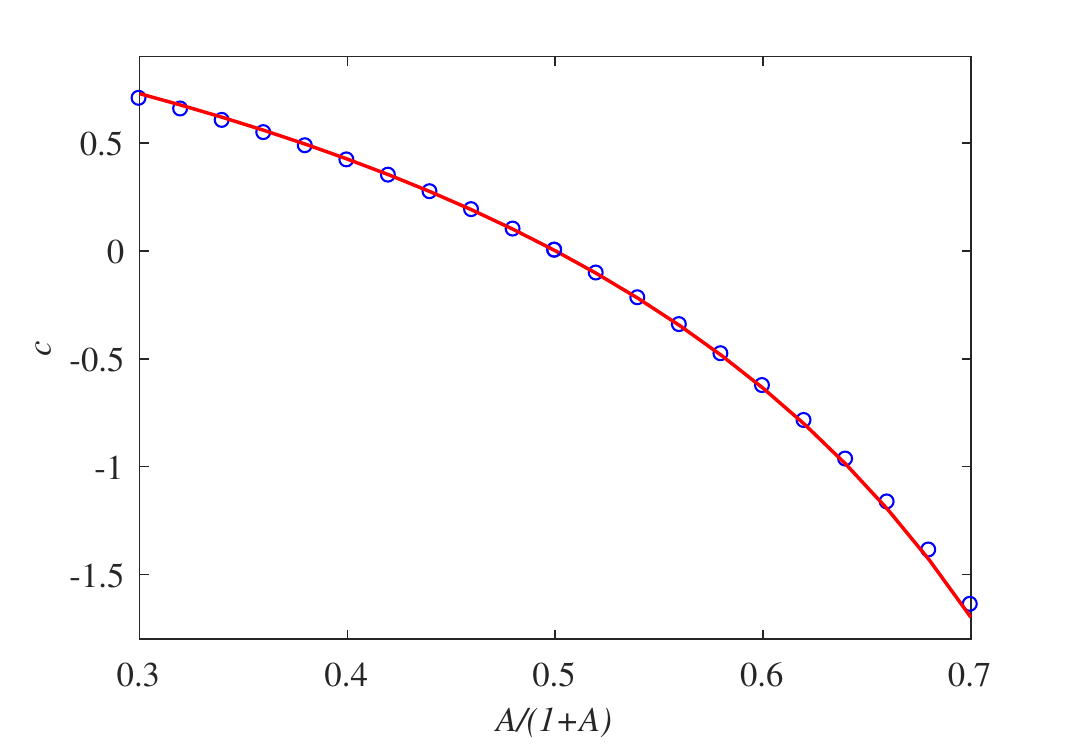}
\caption{(color online) The numerically generated wave speed for the $V\to U$ front. The solid (red) curve corresponds to the analytic prediction, and the (blue) circles are the approximate wave speed derived from a numerical simulation of the PDE \eref{e:4} with $R=8$ using the standard second-order finite difference schemes to approximate the spatial derivatives.}
\label{f:WaveSpeed}
\end{center}
\end{figure}

If $A\neq1$, numerical simulations indicate that the compacton fronts will travel at a constant speed which depends upon $A$.  Moreover, the simulations suggest that the shape of the front at a fixed time is roughly that of the compacton for $A=1$. In order to derive an approximate analytic expression for the wavespeed we plug $U_\rmc(x+ct)$ into the PDE \eref{e:4}, multiply the resultant equation by $\partial_xU_\rmc(x+ct)$, and then integrate over the domain where the front is nonconstant. Doing all this leads to the following predictions for the wave-speed,
\begin{equation}\label{e:cpred}
V\to U,\,\,c=-\frac{\sqrt{2R}}{\pi}(A-1);\quad
U\to V,\,\,c=\frac{\sqrt{2R}}{\pi}(A-1).
\end{equation}
The notation $j\to k$ corresponds to the front which has value $j$ for
$x\ll0$ and value $k$ for  $x\gg0$. See \autoref{f:WaveSpeed} for the comparison of the theoretical prediction with the results of a numerical simulation of the PDE \eref{e:4}. Numerical simulations indicate
that these are good predictions for a relatively large range of $A$
for the 1D PDE model; recall the relevant
discussion also in~\autoref{f:2DFrontTravel}. Moreover, we find that for $R$ sufficiently large, and away from the saddle-node bifurcation points, these are also good predictions for the wave-speed for the discrete model. 

\begin{remark}
If $A<1$, so that language $U$ is preferred, the front will move so that language $U$ invades language $V$. On the other hand, if $A>1$, so that $V$ is preferred, $V$ will invade $U$. The standing compacton which exists for $A=1$ is then seen as a transition between these two invasion fronts.
\end{remark}

\subsection{Spectral stability: one dimension}\label{s:ssod}

Let us now consider the spectral stability of these compactons. 
The 1D version of the PDE \eref{e:42d} is,
\begin{equation}\label{e:4}
	\partial_tu=R u(1-u)\left[(1+A)u-A\right]+(1+A)u(1-u)\partial_x^2u+\left[1-(1+A)u\right](\partial_xu)^2.
\end{equation}
Writing $u=U_\rmc+v$, 
when $A=1$ the linearized problem for $v$ is,
\begin{equation}\label{e:4a}
	\partial_tv=2\partial_x\left[U_\rmc(1-U_\rmc)\partial_xv\right]+g(U_\rmc)v,
\end{equation}
where,
\[
g(U_\rmc)=R(-6U_\rmc^2+6U_\rmc-1)+ 2(1-2U_\rmc)\partial_x^2U_\rmc-2(\partial_xU_\rmc)^2.
\]

Without loss of generality assume the solution in question is the $V\to U$ front, i.e., $U_\rmc(x)=0$ for $x\le-\pi\sqrt{2/R}$, and $U_\rmc(x)=1$ for $x\ge0$. Outside the interval $[-\pi\sqrt{2/R},0]$ the linearized PDE \eref{e:4a} becomes an ODE,
\[
\partial_tv=-Rv.
\]
The associated spectral problem is,
\[
\lambda v=-Rv\quad\leadsto\quad\lambda=-R,\,\,\mathrm{or}\,\,v\equiv0.
\]
Because of the degeneracy associated with the diffusion coefficient, the essential spectrum for the operator comprises a single point. On the other hand, if  then upon using the expression for the compacton the associated spectral problem is the singular Sturm-Liouville problem,
\begin{equation}\label{e:4b}
	\frac12\partial_x\left[\sin^2\left(\sqrt{\frac{R}2}\,x\right)\partial_xv\right]-
	\frac{R}4\left(3\cos^2\left(\sqrt{\frac{R}2}\,x\right)-1\right)v=\lambda v.
\end{equation}
If $\lambda\neq-R$, then for the sake of continuity we need Dirichlet boundary conditions at the endpoints,
\[
v\left(-\sqrt{\frac2{R}}\,\pi\right)=v(0)=0.
\]
Regarding the interior problem, $x\in[-\pi\sqrt{2/R},0]$, due to spatial translation a solution when $\lambda=0$ is $v_0(x)=\partial_xU_\rmc$. Since the front is monotone, this eigenfunction is of one sign. Consequently, by classical Sturmian theory $\lambda=0$ is the largest eigenvalue, so the wave is spectrally stable.

Now consider the concatenation of fronts. Since each front is a compacton, there will be no tail-tail interaction leading to small eigenvalues. Consequently, each front will add another eigenvalue associated with the eigenvalue of the original front. The associated eigenfunction will simply be a spatial translation of the associated eigenfunction. In particular, if there are $N$ fronts, then $\lambda=0$ will be a semi-simple eigenvalue with geometric multiplicity $N$. The multiplicity follows from the fact that each front can be spatially translated without affecting any of the other fronts. 

Suppose we have two fronts, so the solution is a flat-topped compacton. As the size of the top is nonzero, there will be two zero eigenvalues, and the rest of the spectrum will be negative. At the limit of a zero length top we have the pulse compacton,
\[
U_\rmc(x)=\frac12\left[1+\cos\left(\sqrt{\frac{R}2}\,x\right)\right],\quad
-\sqrt{\frac2{R}}\,\pi\le x\le\sqrt{\frac2{R}}\,\pi.
\]
Since the diffusion is zero at $x=0$, so the eigenvalue problem is still degenerate, we can still think of this solution as the concatenation of two fronts, a left front and a right front. The eigenvalue at zero will have geometric multiplicity two. One eigenfunction will be $\partial_xU_\rmc$ of the left front, and zero elsewhere, while another will be $\partial_xU_\rmc$ of the right front, and zero elsewhere. Using linearity, we note that one eigenfunction is the sum of these two, which is precisely the expected spatial translation eigenfunction of the full compacton, $\partial_xU_\rmc$.

\subsection{Spectral stability: two dimensions}

A steady-state front solution to the 2D model \eref{e:4} when $A=1$ is the compacton, $u(x,y)=U_\rmc(x)$. As we saw in \autoref{s:ssod}, for the 1D model \eref{e:4} the original front is spectrally stable with a simple zero eigenvalue, and a concatenation of $N$ fronts is spectrally stable with a semi-simple zero eigenvalue of multiplicity $N$. Let $U(x)$ represent a spectrally stable concatenation of $N$ fronts, which is a stripe pattern.

Consider the spectral stability of the stripes for the full 2D problem. Denote the 1D self-adjoint linearization in \eref{e:4a} about the concatenation as $\calL_1$. The linearization about this striped pattern for \eref{e:42d} is,
\[
\calL_2=\calL_1+2U(1-U)\partial_y^2,
\]
which is also self-adjoint.
Using the Fourier transform to write candidate eigenfunctions,
\[
w(x,y)=v(x)\rme^{\rmi\xi y},
\]
we have,
\[
\calL_2w=\left(\calL_1-2\xi^2U(1-U)\right)v\rme^{\rmi\xi y}.
\]
We already know $\calL_1$ is a nonpositive self-adjoint operator. Since $\xi^2U(1-U)\ge0$, we can therefore conclude $\calL_2$ is a nonpositive self-adjoint operator. Consequently, there are no positive eigenvalues, so the stripe pattern inherits the spectral stability of the concatenation. In particular, it is spectrally stable.

\section{Spots: a case study}

We now consider the existence and spectral stability of spots. A spot is a contiguous set of sites on the lattice which all share language $U$ (or $V$). All other sites share language $V$ (or $U$). For example, a $2\times3$ spot will be a rectangle of height 2 and length 3, so there will be 6 total sites which share language $U$.  When $\epsilon_1=0$ a stable spot of any size and shapecan be formed.  By the Implicit Function Theorem the spot will persist and be spectrally stable for small $\epsilon_1$. Our goal here is to construct a snaking diagram for this spot, and then briefly discuss the dynamics associated with small perturbations of a spot.

\subsection{Existence}

\begin{figure}[ht]
\begin{center}
\begin{tabular}{cc}
\includegraphics{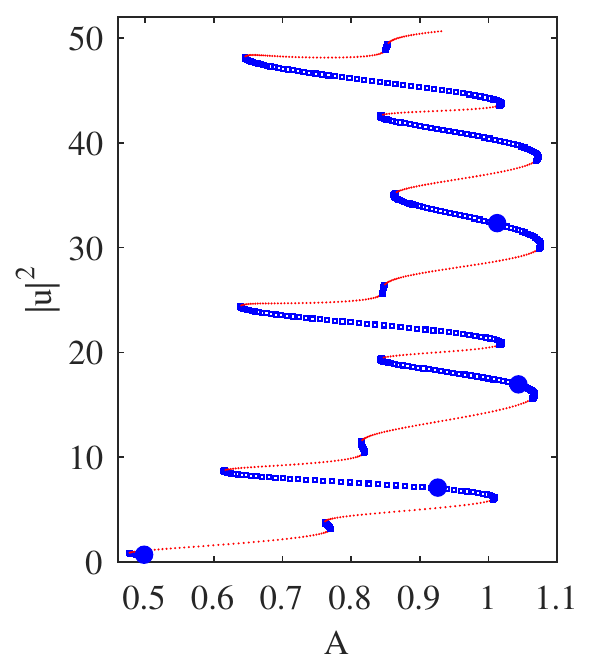}&
\includegraphics[trim = 0 -25 0 0]{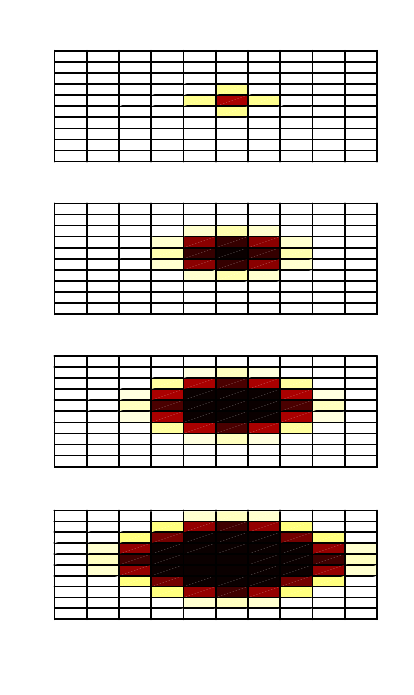}
\end{tabular}
\caption{(color online) The numerically generated snaking diagram for
  a square lattice of size $20\times20$ when $R_1=0.1$ and starting
  with a $1\times1$  spot. The figure on the left is the snaking
  diagram, and the figures on the right provide stable solutions
  arising from the diagram. The notation on the vertical axis, $|u|^2$, represents the square of the $ \ell^2$-norm of the solution. The upper right panel has
  $(A,|u|^2)\sim(0.5,0.6139)$,
  the next one down has $(A,|u|^2)\sim(0.9276,7.0127)$, the third one down
  has $(A,|u|^2)\sim(1.0449,16.8568)$, and the bottom panel on
  the right has $(A,|u|^2)\sim(1.0139,32.2287)$. Each of these points is marked by a large (blue) filled dot on the snaking diagram. For the snaking diagram stable solutions are marked by a (blue) square, and unstable solutions are marked with a (red) dot. While we do not show it here, the growth in terms of the total number of contiguous groups holding language $U$ appears to have no upper bound.}
\label{f:SnakingDiagram1x1-3x1}
\end{center}
\end{figure}

First consider the snaking diagram associated with a steady-state solution. We will start with the configurations at $R_1=0$ of a $1\times1$ square of $U$ sitting on a background of $V$. The results are plotted in \autoref{f:SnakingDiagram1x1-3x1}. The figure on the left gives the snaking diagram, and some stable solutions arising from the snaking are given on the right. For the snaking diagram stable solutions are marked with a (blue) square, and unstable solutions are marked with a (red) dot. The initial $1\times1$ configuration grows seemingly without bound. While we do not provide all the pictures here, as the norm of the solution grows the shape of the contiguous $U$ speakers for a stable solution is either a square or something that has roughly a circular geometry. Regarding the transition from stable to unstable solutions, it is generally not a saddle-node bifurcation, e.g., at the transition point the number of unstable eigenvalues will go from zero to two. Moreover, within the curve of unstable solutions there are additional bifurcations where the number of positive eigenvalues either increases or decreases. The solution structure is rich, but we leave a detailed look at it for a different paper.

\begin{remark}
We should point out that as in the case of single stripes being concatenated to form more complicated stripe patterns, we can concatenate single spots to form more complicated structures.  All that is required for each spot to essentially be an isolated structure is for the spots to be sufficiently separated.  Our experience is that a minimal separation distance between two adjacent spots of three sites is enough.
\end{remark}

\subsection{Dynamics}

\begin{figure}[ht]
\begin{center}
\includegraphics{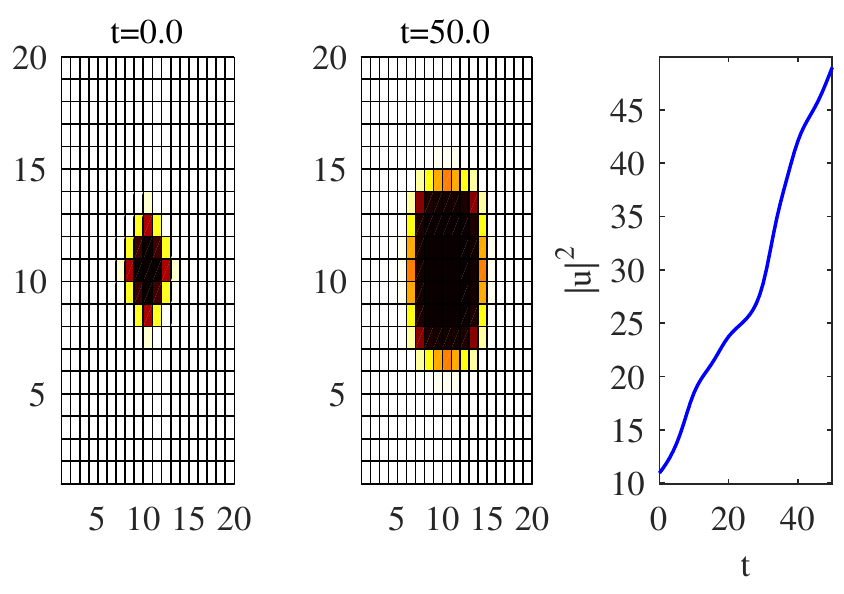}
\caption{(color online) The time evolution of an $A=0.8165$ solution when $A=0.6$. The panel on the far right shows the evolution of the square of the $ \ell^2$-norm of the solution.}
\label{f:TwoClumpEvolveA060}
\end{center}
\end{figure}

Now let us consider the dynamical implications of the snaking diagram.  In particular, we shall look at the effect of varying $A$ for fixed $R_1=0.1$.  Recall that for stripes we saw in \autoref{s:32aa} that outside the snaking diagram traveling waves would appear; in particular, if $A<A_-$ then language $U$ would invade language $V$, whereas if $A>A_+$, then language $V$ would invade language $U$. Consequently, we expect a similar behavior for spots; in particular, a spot will grow or die as a function of the prestige.
For a particular example we start with a stable solution arising from the $1\times1$ initial configuration when $A=0.8165$. The square of the $\ell^2$-norm of this solution is roughly $11$. This solution is contained in the small stable branch shown in \autoref{f:SnakingDiagram1x1-3x1} with $A_-\sim0.8139$ and $A_+\sim0.8192$.

\begin{figure}[ht]
\begin{center}
\includegraphics{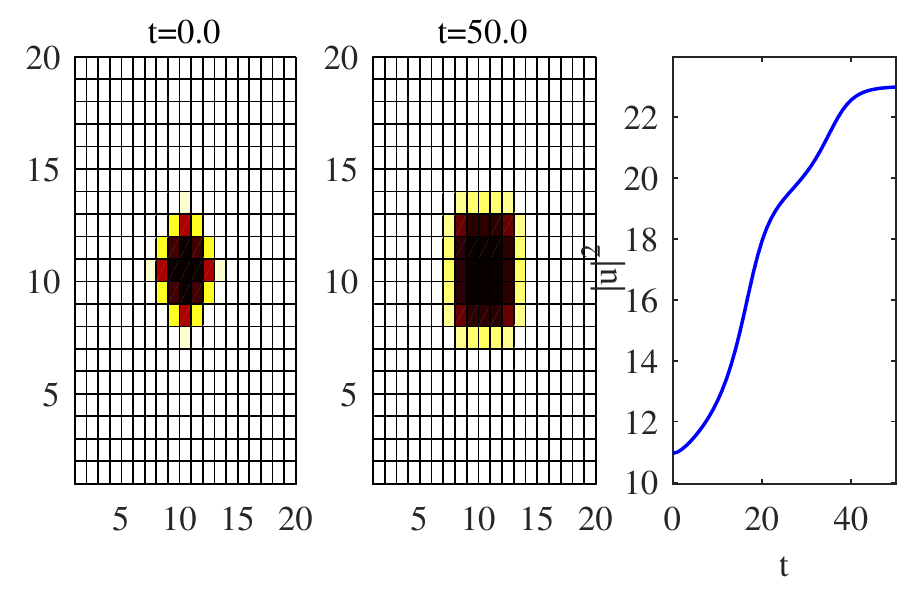}
\caption{(color online) The time evolution of an $A=0.8165$ solution when $A=0.75$. The panel on the far right shows the evolution of the square of the $ \ell^2$-norm of the solution.}
\label{f:TwoClumpEvolveA075}
\end{center}
\end{figure}

First suppose that $A=0.6$. When looking at the snaking diagram, we
see that there are no stable steady-state solutions with this value of
$A$.
The time evolution associated with this initial condition is provided in \autoref{f:TwoClumpEvolveA060}. Of particular interest is the evolution of the square of the norm in the far right panel. We see that the norm is growing up to at least $t=50$. While we do not show it here, the norm continues to grow until the all the nodes share the common language $U$. The growth in language $U$ is manifested in the square becoming larger and larger as those nodes containing $V$ at the boundary between $U$ and $V$ switch to language $U$.

\begin{figure}[ht]
\begin{center}
\includegraphics{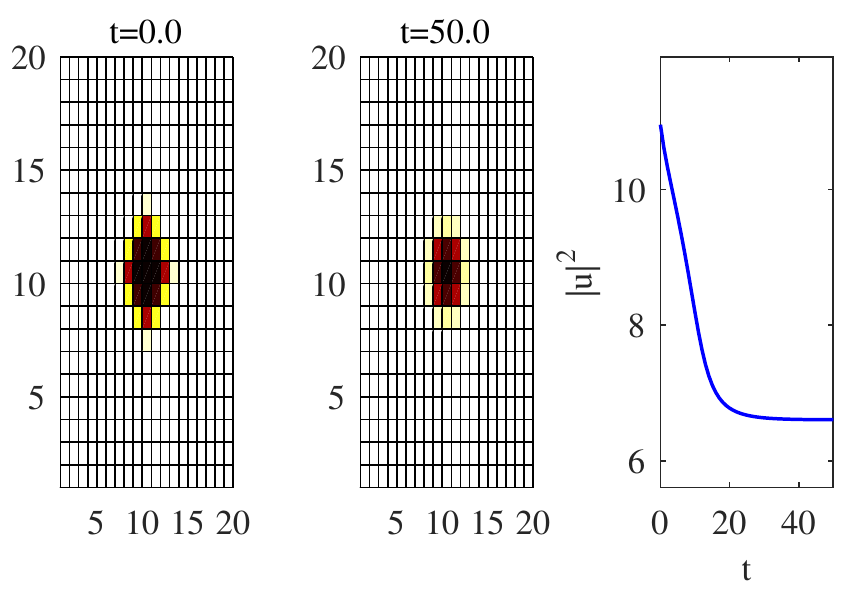}
\caption{(color online) The time evolution of an $A=0.8165$ solution when $A=0.9$. The panel on the far right shows the evolution of the square of the $ \ell^2$-norm of the solution.}
\label{f:TwoClumpEvolveA090}
\end{center}
\end{figure}

Next suppose that $A=0.75$. When looking at the snaking diagram, we
see there is a
(stable) steady-state solution with this value of $A$ and which also
has a larger norm. The time evolution associated with this initial
condition is provided in \autoref{f:TwoClumpEvolveA075}. Of particular
interest is the evolution of the square of the norm in the far right
panel, which in this case achieves a steady-state. The final state at
$t=50$ corresponds to the first stable solution on the snaking diagram
where $a=0.75$, and whose norm is greater than $11$. Language $U$
invades language $V$ until a steady-state  configuration is reached.

\begin{figure}[ht]
\begin{center}
\includegraphics{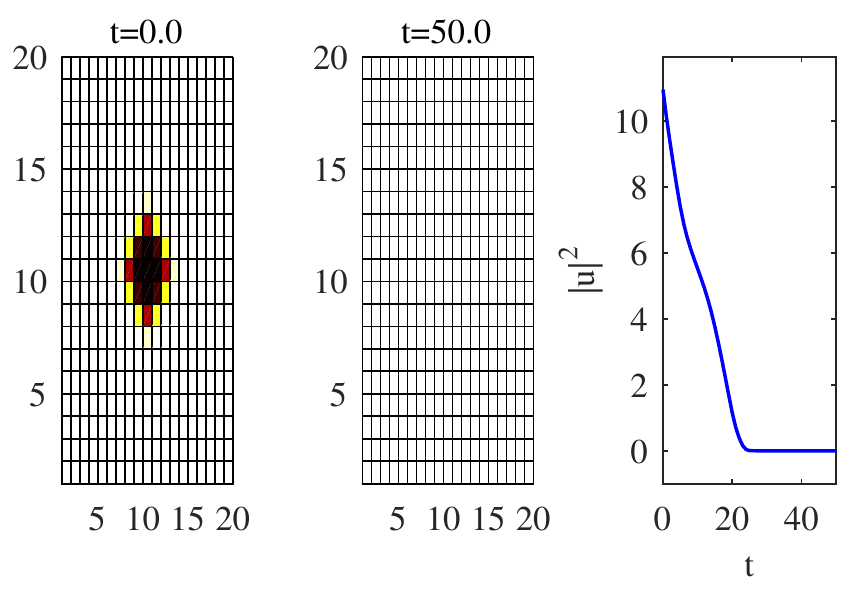}
\caption{(color online) The time evolution of an $A=0.8165$ solution when $A=1.1$. The panel on the far right shows the evolution of the square of the $ \ell^2$-norm of the solution.}
\label{f:TwoClumpEvolveA110}
\end{center}
\end{figure}

For the next example suppose that $A=0.9$.  When looking at the
snaking diagram, we see there is a steady-state solution with this
value of $A$ and which also has a smaller norm. The time evolution
associated with this initial condition is provided in
\autoref{f:TwoClumpEvolveA090}. Of particular interest is the
evolution of the square of the norm in the far right panel, which in
this case also achieves a steady-state. The final state at $t=50$ corresponds to the first stable solution on the snaking diagram where $A=0.9$, and whose norm is less than $11$. Language $V$ invades language $U$ until a steady-state configuration is reached. For the last example suppose that $A=1.1$.
When looking at the snaking diagram, we see there is no steady-state solution with this value of $A$ and which also has a smaller norm. The time evolution associated with this initial condition is provided in \autoref{f:TwoClumpEvolveA110}. Of particular interest is the evolution of the square of the norm in the far right panel, which in this case goes to zero. Language $V$ invades language $U$ until the entire lattice shares the common language $V$.

In conclusion, we have the following rule-of-thumb if the initial
configuration is near a steady state solution. If the value of $A$ is
decreased, so that the prestige of language $U$ increases, then a
 spot of $U$ in a sea of $V$ will grow until a stable steady-state
associated with that value of $A$ is achieved.  If no such
steady-state exists, then eventually the entire lattice will share
language $U$. On the other hand, if the value of $A$ is increased, so
that the prestige of language $V$ increases, then a  spot of $U$ in a
sea of $V$ will shrink in size until a stable steady-state associated
with that value of $A$ is achieved.  If no such steady-state exists,
then eventually the entire lattice will share language $V$. While we
do not show it here, this rule was manifested in every numerical
simulation that we performed. It would be most interesting to
translate this observation into a precise  mathematical
statement. This is left as an interesting direction for future work.

\section{Conclusions \& Future Challenges}

We have derived an ODE model of language dynamics on a square lattice
which is a natural generalization of the AS language model on one
lattice site. The model can also be used to discuss, e.g., the spread
of an opinion through the lattice, or the growth/decay of religious
observance on the lattice. We also looked at the continuum limit of
the ODE, which is a PDE which features a degenerate diffusion term. We
numerically studied the existence of special spatial structures on the
lattice; primarily, stripes and spots. Through a combination of
numerics and analysis we analyzed the dynamics associated with small
perturbations of these spatial structures. Finally, we provided
rules-of-thumb to help understand how languages die and grow in terms
of their prestige, and interaction with neighboring communities.

As is already evident from the discussion above, there are numerous
directions in this emerging field that are worthwhile of further
study.
Some are already concerning the model at hand. As highlighted earlier,
features such as the bifurcation of traveling solutions from standing
ones and their scaling laws, or the more precise identification of
the discrete solutions and their tails from a mathematical analysis
perspective would be of interest. While it is unclear whether
something
analytical can be said about the bifurcation diagram of genuinely
two-dimensional
states such as spots, our numerical observations regarding the model
dynamics formulate a well-defined set of conjectures regarding the
fate of a spot when the prestige is decreased or increased that
may be relevant to further explore mathematically. However, it would
also be relevant to consider variations of the model. Here, we
selected
as a first step of study to explore an ordered two-dimensional square
lattice. However, the $\vI_{jk}$ may be relevant to generalize to more
complex networks and modified (influence or) ``adjacency matrices''
to explore their impact on the findings presented herein.
As indicated herein, the role of near-neighbor interactions is
expected to maintain some of the key features we considered; yet in
a progressively connected world, the consideration of nonlocal,
long-range
interactions may be of interest in its own right.
Another
possibility is to insert a spatially heterogeneous prestige $A_{jk}$
and examine how its spatial variation may influence standing and
traveling structures. There  are numerous variants that can be
considered
thereafter, e.g., how does a local prestige variation interact with the
traveling wave patterns explored herein? Such queries have been
considered in other contexts where the interactions bear
a linear component recently, e.g., see \citet{hoffman:esf17},
but have yet to be considered in a fully nonlinear setting such
as the one herein. Such studies, as applicable, will be reported in future publications.

\begin{thebibliography}{36}
	\providecommand{\natexlab}[1]{#1}
	\providecommand{\url}[1]{\texttt{#1}}
	\expandafter\ifx\csname urlstyle\endcsname\relax
	\providecommand{\doi}[1]{doi: #1}\else
	\providecommand{\doi}{doi: \begingroup \urlstyle{rm}\Url}\fi
	
	\bibitem[Abrams and Strogatz(2003)]{abrams:mtd03}
	D.~Abrams and S.~Strogatz.
	\newblock Modelling the dynamics of language death.
	\newblock \emph{Nature}, 424:\penalty0 900, 2003.
	
	\bibitem[Abrams et~al.(2011)Abrams, Yaple, and Wiener]{abrams:dos11}
	D.~Abrams, H.~Yaple, and R.~Wiener.
	\newblock Dynamics of social group competition: modeling the decline of
	religious affiliation.
	\newblock \emph{Phys. Rev. Lett.}, 107:\penalty0 088701, 2011.
	
	\bibitem[Amano et~al.(2014)Amano, Sandel, Eager, Bulteau, Svenning, Dalsgaard,
	Rahbek, Davies, and Sutherland]{amano:gda14}
	T.~Amano, B.~Sandel, H.~Eager, E.~Bulteau, J.-C. Svenning, B.~Dalsgaard,
	C.~Rahbek, R.~Davies, and W.~Sutherland.
	\newblock Global distribution and drivers of language extinction risk.
	\newblock \emph{Proc. Roy. Society B: Biological Sciences}, 281:\penalty0
	20141574, 2014.
	
	\bibitem[Anderson et~al.(2016)Anderson, Faye, Scheel, and
	Stauffer]{anderson:pau16}
	T.~Anderson, G.~Faye, A.~Scheel, and D.~Stauffer.
	\newblock Pinning and unpinning in nonlocal systems.
	\newblock \emph{J. Dyn. Diff. Eq.}, 28:\penalty0 897--923, 2016.
	
	\bibitem[Bramburger and Sandstede(2020{\natexlab{a}})]{bramburger:lpi20}
	J.~Bramburger and B.~Sandstede.
	\newblock Localized patterns in planar bistable weakly coupled lattice systems.
	\newblock \emph{Nonlinearity}, 33:\penalty0 3500--3525, 2020{\natexlab{a}}.
	
	\bibitem[Bramburger and Sandstede(2020{\natexlab{b}})]{bramburger:sls20}
	J.~Bramburger and B.~Sandstede.
	\newblock Spatially localized structures in lattice dynamical systems.
	\newblock \emph{J. Nonlinear Science}, 30:\penalty0 603--644,
	2020{\natexlab{b}}.
	
	\bibitem[Bujalski et~al.(2018)Bujalski, Dwyer, Kapitula, Le, Malvai,
	Rosental-Kay, and Ruiter]{bujalski:cac18}
	J.~Bujalski, G.~Dwyer, T.~Kapitula, Q.-N. Le, H.~Malvai, J.~Rosental-Kay, and
	J.~Ruiter.
	\newblock Consensus and clustering in opinion formation on networks.
	\newblock \emph{Phil. Trans. R. Soc. A}, 376:\penalty0 20170186, 2018.
	
	\bibitem[Colucci et~al.(2014)Colucci, Mira, Nieto, and
	Otero-Espinar]{colucci:cie14}
	R.~Colucci, J.~Mira, J.~Nieto, and M.~Otero-Espinar.
	\newblock Coexistence in exotic scenarios of a modified {A}brams-{S}trogatz
	model.
	\newblock \emph{Complexity}, 21\penalty0 (4):\penalty0 86--93, 2014.
	
	\bibitem[Dhooge et~al.(2003)Dhooge, Govaerts, and Kuznetsov]{dhooge:mam03}
	A.~Dhooge, W.~Govaerts, and Y.~Kuznetsov.
	\newblock Matcont: a {MATLAB} package for numerical bifurcation analysis of
	{ODE}s.
	\newblock \emph{ACM TOMS}, 29:\penalty0 141--164, 2003.
	
	\bibitem[Eekhoff(2019)]{eekhoff:ofd19}
	K.~Eekhoff.
	\newblock Opinion formation dynamics with contrarians and zealots.
	\newblock \emph{SIAM J. Undergraduate Research Online}, 12, 2019.
	
	\bibitem[Elmer and Vleck(2002)]{elmer:avo02}
	C.~Elmer and E.~Van Vleck.
	\newblock A variant of {N}ewton's method for the computation of traveling waves
	of bistable differential-difference equations.
	\newblock \emph{J. Dyn. Diff. Eq.}, 14\penalty0 (3):\penalty0 493--517, 2002.
	
	\bibitem[Fujie et~al.(2013)Fujie, Aihara, and Masuda]{fujie:amo13}
	R.~Fujie, K.~Aihara, and N.~Masuda.
	\newblock A model of competition among more than two languages.
	\newblock \emph{J. Stat. Phys.}, 151:\penalty0 289--303, 2013.
	
	\bibitem[Haragus and Scheel(2006)]{haragus:cdi06}
	M.~Haragus and A.~Scheel.
	\newblock Corner defects in almost planar interface propagation.
	\newblock \emph{Ann. Inst. H. Poincare (C) Anal. Non Lineaire}, 23:\penalty0
	283--329, 2006.
	
	\bibitem[Hoffman et~al.(2017)Hoffman, Hupkes, and Vleck]{hoffman:esf17}
	A.~Hoffman, H.~Hupkes, and E.~Van Vleck.
	\newblock \emph{Entire Solutions for Bistable Lattice Differential Equations
		with Obstacles}, volume 250 of \emph{Memoirs Am. Math. Soc.}
	\newblock Am. Math. Soc., 2017.
	
	\bibitem[Hupkes and Lunel(2005)]{hupkes:aon05}
	H.~Hupkes and S.~Verdun Lunel.
	\newblock Analysis of {N}ewton's method to compute travelling waves in discrete
	media.
	\newblock \emph{J. Dyn. Diff. Eq.}, 17\penalty0 (3):\penalty0 523--572, 2005.
	
	\bibitem[Hupkes and Sandstede(2013)]{hupkes:sop13}
	H.~Hupkes and B.~Sandstede.
	\newblock Stability of pulse solutions for the discrete fitzhugh-nagumo system.
	\newblock \emph{Trans. Amer. Math. Soc.}, 365:\penalty0 251--301, 2013.
	
	\bibitem[Hupkes et~al.(2011)Hupkes, Pelinovsky, and Sandstede]{hupkes:pfi11}
	H.~Hupkes, D.~Pelinovsky, and B.~Sandstede.
	\newblock Propagation failure in the discrete {N}agumo equation.
	\newblock \emph{Proc. Amer. Math. Soc.}, 139\penalty0 (10):\penalty0
	3537--3551, 2011.
	
	\bibitem[Juane et~al.(2019)Juane, Seoane, nuzuri, and Mira]{juane:uat19}
	M.~Juane, L.~Seoane, A.~Mu\ nuzuri, and J.~Mira.
	\newblock Urbanity and the dynamics of language shift in {G}alicia.
	\newblock \emph{Nature Comm.}, 10:\penalty0 1680, 2019.
	
	\bibitem[Kapitula(1997)]{kapitula:mso97}
	T.~Kapitula.
	\newblock Multidimensional stability of planar travelling waves.
	\newblock \emph{Trans. AMS}, 349\penalty0 (1):\penalty0 257--269, 1997.
	
	\bibitem[Kevrekidis et~al.(2001)Kevrekidis, Kevrekidis, and
	Bishop]{kevrekidis:pfu01}
	P.~Kevrekidis, I.~Kevrekidis, and A.~Bishop.
	\newblock Propagation failure, universal scalings and {G}oldstone modes.
	\newblock \emph{Phys. Lett. A}, 279\penalty0 (5-6):\penalty0 361--369, 2001.
	
	\bibitem[Marvel et~al.(2012)Marvel, Hong, Papush, and Strogatz]{marvel:emc12}
	S.~Marvel, H.~Hong, A.~Papush, and S.~Strogatz.
	\newblock Encouraging moderation: clues from a simple model of ideological
	conflict.
	\newblock \emph{Phys. Rev. Lett.}, 109:\penalty0 118702, 2012.
	
	\bibitem[Mira and Paredes(2005)]{mira:isa05}
	J.~Mira and \'A. Paredes.
	\newblock Interlinguistic similarity and language death dynamics.
	\newblock \emph{Europhys. Lett.}, 69\penalty0 (6):\penalty0 1031--1034, 2005.
	
	\bibitem[Mira et~al.(2011)Mira, Seoane, and Nieto]{mira:tio11}
	J.~Mira, L.~Seoane, and J.~Nieto.
	\newblock The importance of interlinguistic similarity and stable bilingualism
	when two languages compete.
	\newblock \emph{New J. Physics}, 13:\penalty0 033007, 2011.
	
	\bibitem[Moore and Promislow(2005)]{moore:rgr05}
	R.~Moore and K.~Promislow.
	\newblock Renormalization group reduction of pulse dynamics in thermally loaded
	optical parametric oscillators.
	\newblock \emph{Physica D}, 206:\penalty0 62--81, 2005.
	
	\bibitem[Otero-Espinar et~al.(2013)Otero-Espinar, Seoane, Nieto, and
	Mira]{otero:aas13}
	M.~Otero-Espinar, L.~Seoane, J.~Nieto, and J.~Mira.
	\newblock An analytic solution of a model of language competition with
	bilingualism and interlinguistic similarity.
	\newblock \emph{Physica D}, 264:\penalty0 17--26, 2013.
	
	\bibitem[Parker et~al.(2020)Parker, Kevrekidis, and Sandstede]{parker:eas20}
	R.~Parker, P.~Kevrekidis, and B.~Sandstede.
	\newblock Existence and spectral stability of multi-pulses in discrete
	{H}amiltonian lattice systems.
	\newblock \emph{Physica D}, 408:\penalty0 132414, 2020.
	
	\bibitem[Promislow(2002)]{promislow:arm02}
	K.~Promislow.
	\newblock A renormalization method for modulational stability of quasi-steady
	patterns in dispersive systems.
	\newblock \emph{SIAM J. Math. Anal.}, 33\penalty0 (6):\penalty0 1455--1482,
	2002.
	
	\bibitem[Sandstede(1998)]{sandstede:som98}
	B.~Sandstede.
	\newblock Stability of multiple-pulse solutions.
	\newblock \emph{Trans. Amer. Math. Soc.}, 350:\penalty0 429--472, 1998.
	
	\bibitem[Sandstede(2002)]{sandstede:sot02}
	B.~Sandstede.
	\newblock Stability of travelling waves.
	\newblock In \emph{Handbook of Dynamical Systems}, volume~2, chapter~18, pages
	983--1055. Elsevier Science, 2002.
	
	\bibitem[Stauffer et~al.(2007)Stauffer, Castell\'o, Egu\'iluz, and
	Miguel]{stauffer:mas07}
	D.~Stauffer, X.~Castell\'o, V.~Egu\'iluz, and M.~San Miguel.
	\newblock Microscopic {A}brams-{S}trogatz model of language competition.
	\newblock \emph{Physica A}, 374:\penalty0 835--842, 2007.
	
	\bibitem[Tanabe and Masuda(2013)]{tanabe:cdo13}
	S.~Tanabe and N.~Masuda.
	\newblock Complex dynamics of a nonlinear voter model with contrarian agents.
	\newblock \emph{Chaos}, 23:\penalty0 043136, 2013.
	
	\bibitem[Vazquez et~al.(2010)Vazquez, Castell\'o, and Miguel]{vazquez:abm10}
	F.~Vazquez, X.~Castell\'o, and M.~San Miguel.
	\newblock Agent based models of language competition: macroscopic descriptions
	and order-disorder transitions.
	\newblock \emph{J. Stat. Mech.}, page P04007, 2010.
	
	\bibitem[Vidal-Franco et~al.(2017)Vidal-Franco, Guiu-Souto, and
	nuzuri]{franco:sme17}
	I.~Vidal-Franco, J.~Guiu-Souto, and A.~Mu\ nuzuri.
	\newblock Social media enhances languages differentiation: a mathematical
	description.
	\newblock \emph{Royal Society Open Science}, 4:\penalty0 170094, 2017.
	
	\bibitem[Wang et~al.(2016)Wang, Rong, and Wu]{wang:bam16}
	S.~Wang, L.~Rong, and J.~Wu.
	\newblock Bistability and multistability in opinion dynamics models.
	\newblock \emph{Applied Math. Comp.}, 289:\penalty0 388--395, 2016.
	
	\bibitem[Yun et~al.(2016)Yun, Shang, Wei, Liu, and Li]{yun:tpo16}
	J.~Yun, S.-C. Shang, X.-D. Wei, S.~Liu, and Z.-J. Li.
	\newblock The possibility of coexistence and co-development in language
	competition: ecology-society computational model and simulation.
	\newblock \emph{SpringerPlus}, 5:\penalty0 855, 2016.
	
	\bibitem[Zhou et~al.(2019)Zhou, Szymanski, and Gao]{zhou:mce19}
	Z.~Zhou, B.~Szymanski, and J.~Gao.
	\newblock Modeling competitive evolution of multiple languages.
	\newblock arXiv:1907.06848v1, 2019.
	
\end{thebibliography}
\bibliographystyle{plainnat}

\end{document}